\documentclass{article}

\usepackage[all,knot]{xy}
\xyoption{arc}
\usepackage{latexsym,graphics,epsfig}
\usepackage{rotate}
\usepackage{amsfonts,amssymb,amsthm}
\usepackage{hyperref}

\newcommand{\R}{{\mathbb R}}

\newcommand{\Z}{{\mathbb Z}}

\newcommand{\maps}{\colon}
\newcommand{\iso}{\cong}

\def\stackto #1 { \, {\stackrel{#1}{\longrightarrow}}\, }
\def\stackTo #1 { {\stackrel{#1}{\Longrightarrow}} }
\newcommand{\To}{\Rightarrow}

\newcommand{\dt}{\underline{t}}
\newcommand{\dalpha}{\underline{\alpha}}

\newcommand{\U}{{\rm U}}

\newcommand{\sstring}{\mathfrak{string}}
\newcommand{\sugra}{\mathfrak{sugra}}
\newcommand{\SO}{{\rm SO}}
\newcommand{\so}{\mathfrak{so}}

\newcommand{\Aut}{{\rm Aut}}
\newcommand{\OUT}{\mathcal{OUT}}
\newcommand{\AUT}{\mathcal{AUT}}
\newcommand{\aut}{\mathfrak{aut}}
\newcommand{\ZZ}{\mathcal{Z}}
\newcommand{\CC}{\mathcal{C}}
\newcommand{\STRING}{{\cal{STRING}}}

\newcommand{\T}{\mathcal{T}}
\newcommand{\INN}{\mathcal{INN}}

\newcommand{\g}{\mathfrak{g}}
\newcommand{\h}{\mathfrak{h}}

\newcommand{\Set}{\mathrm{Set}}

\newcommand{\Lie}{\mathrm{Lie}}

\newcommand{\G}{\mathcal{G}}
\renewcommand{\T}{\mathcal{T}}

\renewcommand{\P}{\mathcal{P}}

\newcommand{\hol}{{\rm hol}}
\newcommand{\poexp}{\mathcal{P}\exp}

\newcommand{\half}{\frac{1}{2}}

\newtheorem{thm}{Theorem}    
\newtheorem*{unnumbered-thm}{Theorem}    

\newtheorem*{lem}{Lemma}

\theoremstyle{definition}

\newcommand{\define}[1]{{\bf #1}}
\newcommand{\tr}{{\mathrm{tr}}} 

\newcommand{\hepth}[1]{\href{http://arxiv.org/abs/hep-th/#1}{arXiv:hep-th/#1}}

\newcommand{\grqc}[1]{\href{http://arxiv.org/abs/gr-qc/#1}{arXiv:gr-qc/#1}}

\newcommand{\qalg}[1]{\href{http://arxiv.org/abs/q-alg/#1}{arXiv:q-alg/#1}}

\newcommand{\Math}[2]{\href{http://arxiv.org/abs/math.#1/#2}{arXiv:math.#1/#2}}

\newcommand{\MMath}[1]{\href{http://arxiv.org/abs/math/#1}{arXiv:math/#1}}

\newcommand{\arxiv}[1]{\href{http://arxiv.org/abs/#1/}{arXiv:#1}}

\begin{document}

\title{An Invitation to Higher Gauge Theory}

\author{John C.\ Baez and John Huerta\\
Department of Mathematics,  University of California\\
Riverside, California 92521 \\
USA \\
\\
email: baez@math.ucr.edu \qquad huerta@math.ucr.edu}

\date{March 24, 2010}

\maketitle

\begin{abstract}
\noindent
In this easy introduction to higher gauge theory, we describe parallel
transport for particles and strings in terms of 2-connections on
2-bundles.  Just as ordinary gauge theory involves a gauge group, this
generalization involves a gauge `2-group'.  We focus on 6 examples.
First, every abelian Lie group gives a Lie 2-group; the case of
$\U(1)$ yields the theory of $\U(1)$ gerbes, which play an important
role in string theory and multisymplectic geometry.  Second, every
group representation gives a Lie 2-group; the representation of the
Lorentz group on 4d Minkowski spacetime gives the Poincar\'e 2-group,
which leads to a spin foam model for Minkowski spacetime.  Third,
taking the adjoint representation of any Lie group on its own Lie
algebra gives a `tangent 2-group', which serves as a gauge 2-group in
4d $BF$ theory, which has topological gravity as a special
case.  Fourth, every Lie group has an `inner automorphism 2-group',
which serves as the gauge group in 4d $BF$ theory with cosmological
constant term.  Fifth, every Lie group has an `automorphism 2-group',
which plays an important role in the theory of nonabelian gerbes.  And
sixth, every compact simple Lie group gives a `string 2-group'.  We
also touch upon higher structures such as the `gravity 3-group', and
the Lie 3-superalgebra that governs 11-dimensional supergravity.
\end{abstract}


\section{Introduction}
\label{introduction}

Higher gauge theory is a generalization of gauge theory that describes
parallel transport, not just for point particles, but also for
higher-dimensional extended objects.  It is a beautiful new branch of
mathematics, with a lot of room left for exploration.  It has already
been applied to string theory and loop quantum gravity---or more
specifically, spin foam models.  This should not be surprising, since
while these rival approaches to quantum gravity disagree about almost
everything, they both agree that point particles are not enough: we
need higher-dimensional extended objects to build a theory
sufficiently rich to describe the quantum geometry of spacetime.
Indeed, many existing ideas from string theory and supergravity have
recently been clarified by higher gauge theory \cite{Sati,SSS}.  But
we may also hope for applications of higher gauge theory to other less
speculative branches of physics, such as condensed matter physics.

Of course, for this to happen, more physicists need to learn higher
gauge theory.  It would be great to have a comprehensive introduction
to the subject which started from scratch and led the reader to the
frontiers of knowledge.  Unfortunately, mathematical work in this
subject uses a wide array of tools, such as $n$-categories, stacks,
gerbes, Deligne cohomology, $L_\infty$ algebras, Kan complexes, and
$(\infty,1)$-categories, to name just a few.  While these tools are
beautiful, important in their own right, and perhaps necessary for a
deep understanding of higher gauge theory, learning them takes
time---and explaining them all would be a major project.

Our goal here is far more modest.  We shall sketch how to generalize
the theory of parallel transport from point particles to 1-dimensional
objects, such as strings.  We shall do this starting with a bare
minimum of prerequisites: manifolds, differential forms, Lie groups,
Lie algebras, and the traditional theory of parallel transport in
terms of bundles and connections.  We shall give a small taste of the
applications to physics, and point the reader to the literature for
more details.

In Section \ref{connections} we start by explaining categories,
functors, and how parallel transport for particles can be seen as a
functor taking any path in a manifold to the operation of parallel
transport along that path.  In Section \ref{2-connections} we `add
one' and explain how parallel transport for particles and strings can
be seen as `2-functor' between `2-categories'.  This requires that we
generalize Lie groups to `Lie 2-groups'.  In Section \ref{examples}
we describe many examples of Lie 2-groups, and sketch some of their
applications:
\begin{itemize}
\item Section \ref{abelian}: shifted abelian groups, $\U(1)$ gerbes,
and their role in string theory and multisymplectic geometry.
\item Section \ref{poincare}: the Poincar\'e 2-group and the 
spin foam model for 4d Minkowski spacetime.
\item Section \ref{tangent}: tangent 2-groups, 4d $BF$ theory and
topological gravity.  
\item Section \ref{inner}: inner automorphism 2-groups and 
4d $BF$ theory with cosmological constant term.
\item Section \ref{automorphism}: automorphism 2-groups, nonabelian
gerbes, and the gravity 3-group.
\item Section \ref{string}: string 2-groups, string structures, 
the passage from Lie $n$-algebras to Lie $n$-groups, and the
Lie 3-superalgebra governing 11-dimensional supergravity.
\end{itemize}
Finally, in Section \ref{last} we discuss gauge transformations,
curvature and nontrivial 2-bundles.

\section{Categories and Connections}
\label{connections}

A category consists of objects, which we draw as dots:
\[ \bullet \,\, x \]
and morphisms between objects, which we draw as arrows between dots:
\[
\xymatrix{
   x \bullet \ar@/^1pc/[rr]^{f}
&& \bullet y
}
\]
You should think of objects as `things' and morphisms as
`processes'.  The main thing you can do in a category is 
take a morphism from $x$ to $y$ and a morphism from $y$ to $z$:
\[
\xymatrix{
   x \bullet \ar@/^1pc/[rr]^{f}
&& \bullet y \ar@/^1pc/[rr]^{g}
    && \bullet z
}
\]
and `compose' them to get a morphism from $x$ to $z$:
\[
\xymatrix{
   x \bullet \ar@/^2pc/[rrrr]^{gf}
&&&& \bullet z
}
\]

The most famous example is the category $\Set$, which has sets as
objects and functions as morphisms.  Most of us know how to compose
functions, and we have a pretty good intuition of how this works.  So,
it can be helpful to think of morphisms as being like functions.  But
as we shall soon see, there are some very important categories where the
morphisms are \emph{not} functions.

Let us give the formal definition.  A \define{category} consists of:
\begin{itemize} 
\item A collection of \define{objects}, and
\item for any pair of objects $x,y$, a set of \define{morphisms}
$f \maps x \to y$.   Given a morphism $f \maps x \to y$,
we call $x$ its \define{source} and $y$ its \define{target}.
\item Given two morphisms $f \maps x \to y$ and $g \maps y \to z$, 
there is a \define{composite} morphism $gf \maps x \to z$.  Composition 
satisfies the \define{associative law}:
\[ (hg)f = h(gf). \]
\item For any object $x$, there is an \define{identity} morphism
$1_x \maps x \to x$.  These identity morphisms satisfy
the \define{left and right unit laws}:
\[ 1_y f = f = f 1_x \]
for any morphism $f \maps x \to y$.
\end{itemize}
The hardest thing about category theory is getting your arrows to
point the right way.  It is standard in mathematics to use $f g$ to
denote the result of doing first $g$ and then $f$.  In pictures, this
backwards convention can be annoying.  But rather than trying to fight it,
let us give in and draw a morphism $f \maps x \to y$ as an arrow from
right to left:
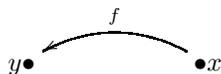
\[
\xymatrix{
   y \bullet 
&& \bullet x \ar@/_1pc/[ll]_{f}
}
\]
Then composition looks a bit better:
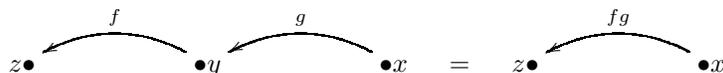
\[
\xymatrix{
   z \bullet 
&& \bullet y \ar@/_1pc/[ll]_{f}
    && \bullet x \ar@/_1pc/[ll]_{g}
}
\quad = \quad
\xymatrix{
   z \bullet 
&& \bullet x \ar@/_1pc/[ll]_{fg}
}
\]

An important example of a category is the `path groupoid' of a space $X$.  
We give the precise definition below, but the basic idea is to 
take the diagrams we have been drawing seriously!  The objects are 
points in $X$, and morphisms are paths: 
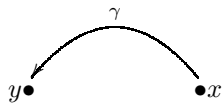
\[
\xymatrix{
   y \bullet 
&& \bullet x \ar@/_2pc/[ll]_{\gamma}
}
\]
We also get examples from groups.  A group is the same as
a category with one object where all the morphisms are invertible.
The morphisms of this category are the elements of the group.  
The object is there just to provide them with a source and target.
We compose the morphisms using the multiplication in the group.

In both these examples, a morphism $f \maps x \to y$ is \textit{not} a
function from $x$ to $y$.  And these two examples have something else
in common: they are important in gauge theory!  We can use a path
groupoid to describe the possible motions of a particle through
spacetime.  We can use a group to describe the symmetries of a
particle.  And when we combine these two examples, we get the concept
of \textit{connection}---the basic field in any gauge theory.

How do we combine these examples?  We do it using a map between
categories.  A map between categories is called a `functor'.  A
functor from a path groupoid to a group will send every object of the
path groupoid to the same object of our group.  After all, a group,
regarded as a category, has only one object.  But this functor will
also send any morphism in our path groupoid to a group element.  In
other words, it will assign a group element to each path in our space.
This group element \emph{describes how a particle transforms} as it
moves along that path.

But this is precisely what a connection does!  A connection lets us
compute for any path a group element describing parallel transport
along that path. So, the language of categories and functors quickly
leads us to the concept of connection---but with an emphasis on
parallel transport.

The following theorem makes these ideas precise.  Let us first
state the theorem, then define the terms involved, and then give
some idea of how it is proved:

\begin{thm} 
\label{thm:connections}
For any Lie group $G$ and any smooth manifold $M$,
there is a one-to-one correspondence between:
\begin{enumerate}
\item connections on the trivial principal $G$-bundle over $M$,
\item $\g$-valued 1-forms on $M$, where $\g$ is the Lie algebra of
$G$, and
\item smooth functors
\[ \hol \maps \P_1(M) \to G \]
where $\P_1(M)$ is the path groupoid of $M$.
\end{enumerate}
\end{thm}
\noindent
We assume you are familiar with the first two items.  
Our goal is to explain the third.  We must start by explaining
the path groupoid.

Suppose $M$ is a manifold.  Then the path groupoid $\P_1(M)$ is
roughly a category in which objects are points of $M$ and a morphism
from $x$ to $y$ is a path from $x$ to $y$.  We compose paths by gluing
them end to end.  So, given a path $\delta$ from $x$ to $y$, and a path
$\gamma$ from $y$ to $z$:
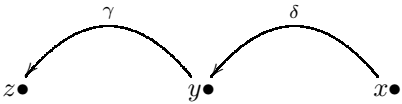
\[
\xymatrix{
   z \bullet 
&& y \bullet \ar@/_2pc/[ll]_{\gamma}
&& x \bullet \ar@/_2pc/[ll]_{\delta}
}
\]
we would like $\gamma \delta$ to be the path from $x$ to $z$ 
built from $\gamma$ and $\delta$.

However, we need to be careful about the details to make sure that the
composite path $\gamma \delta$ is well-defined, and that composition is
associative!  Since we are studying paths in a smooth manifold, we
want them to be smooth.  But the path $\gamma \delta$ may not be
smooth: there could be a `kink' at the point $y$.

There are different ways to get around this problem.  One is to work
with piecewise smooth paths.  But here is another approach: say that a
path
\[  \gamma \maps [0,1] \to M  \]
is {\bf lazy} if it is smooth and also constant in a neighborhood of
$t = 0$ and $t = 1$.  The idea is that a lazy hiker takes a rest
before starting a hike, and also after completing it.  Suppose
$\gamma$ and $\delta$ are smooth paths and $\gamma$ starts where
$\delta$ ends. Then we define their
{\bf composite}
\[   \gamma \delta \maps [0,1] \to M  \]
in the usual way:
\[   (\gamma \delta)(t) = \left\{ \begin{array}{cl}
                           \delta(2t) &\textrm{if } 0 \le t \le \half \\
                           \gamma(2t-1) &\textrm{if } \half \le t \le 1 
\end{array} \right.
\]
In other words, $\gamma \delta$ spends the first half of its time 
moving along $\delta$, and the second half moving along $\gamma$.
In general the path $\gamma \delta$ may not be smooth at $t = \half$.  
However, if $\gamma$ and $\delta$ are lazy, then their composite is
smooth---and it, too, is lazy!  

So, lazy paths are closed under composition.  Unfortunately, 
composition of lazy paths is not associative.  The paths $(\alpha \beta)
\gamma$ and $\alpha (\beta \gamma)$ differ by a smooth
reparametrization, but they are not equal.  To solve this problem, we
can take certain \textit{equivalence classes} of lazy paths as
morphisms in the path groupoid.

We might try `homotopy classes' of paths.  Remember, a homotopy is a
way of interpolating between paths:
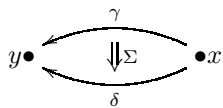
\[
  \xymatrix{
y \bullet 
&& \bullet x
\ar@/^2.5ex/[ll]^{\delta}="g1"\ar@/_2.5ex/[ll]_{\gamma}="g2"
\ar@{=>}^{\Sigma} "g2"+<0ex,-2.5ex>;"g1"+<0ex,2.5ex>
}
\]
More precisely, a \define{homotopy} from the path $\gamma \maps [0,1] \to M$ 
to the path $\delta \maps [0,1] \to M$ is a smooth map
\[ \Sigma \maps [0,1]^2 \to M \]
such that $\Sigma(0,t) = \gamma(t)$ and $\Sigma(1,t) = \delta(t)$.  We
say two paths are \define{homotopic}, or lie in the same
\define{homotopy class}, if there is a homotopy between them.  

There is a well-defined category where the morphisms are homotopy
classes of lazy paths.  Unfortunately this is not right for gauge
theory, since for most connections, parallel transport along homotopic
paths gives different results.  In fact, parallel transport gives the
same result for all homotopic paths if and only if the connection is
\emph{flat}.

So, unless we are willing to settle for flat connections, we need 
a more delicate equivalence relation between paths.  Here the
concept of `thin' homotopy comes to our rescue.  A homotopy is {\bf thin} if
it sweeps out a surface that has zero area.  In other words, it is a
homotopy $\Sigma$ such that the rank of the differential $d\Sigma$ is
less than $2$ at every point.  If two paths differ by a smooth
reparametrization, they are thinly homotopic.  But there are other
examples, too.  For example, suppose we have a path $\gamma \maps x
\to y$, and let $\gamma^{-1} \maps y \to x$ be the \define{reverse}
path, defined as follows:
\[   \gamma^{-1}(t) = \gamma(1-t)  .\]
Then the composite path $\gamma^{-1} \gamma$, which goes from $x$ to
itself:
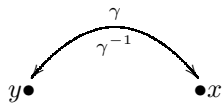
\[  \xymatrix{
y \bullet &&
\bullet x
\ar@/_2pc/@{<->}[ll]^{\gamma^{-1}}_{\gamma} 
}
\]
is thinly homotopic to the constant path that sits at $x$.  
The reason is that we can shrink $\gamma^{-1} \gamma$ down to the 
constant path without sweeping out any area.

We define the \define{path groupoid} $\P_1(M)$ to be the category where:
\begin{itemize}
\item Objects are points of $M$.
\item Morphisms are thin homotopy classes of lazy paths in $M$.
\item If we write $[\gamma]$ to denote the thin homotopy class of the
path $\gamma$, composition is defined by
\[      [\gamma][\delta] = [\gamma \delta] . \]
\item For any point $x \in M$, the identity $1_x$ is the thin homotopy
class of the constant path at $x$. 
\end{itemize}
With these rules, it is easy to check that $\P_1(M)$ is a category.
The most important point is that since the composite paths $(\alpha
\beta) \gamma$ and $\alpha (\beta \gamma)$ differ by a smooth
reparametrization, they are thinly homotopic.  This gives the associative
law when we work with thin homotopy classes.

But as its name suggests, $\P_1(M)$ is better than a mere category.
It is a \define{groupoid}: that is, a category where every morphism
$\gamma \maps x \to y$ has an \define{inverse} $\gamma^{-1} \maps y
\to x$ satisfying
\[  \gamma^{-1} \gamma = 1_x \quad \textrm{and} \quad
 \gamma \gamma^{-1} = 1_y \]
In $\P_1(M)$, the inverse is defined using the concept of a reverse
path:
\[         [\gamma]^{-1} = [\gamma^{-1}]  .\]
The rules for an inverse only hold in $\P_1(M)$ \emph{after} we take
thin homotopy classes.  After all, the composites $\gamma \gamma^{-1}$
and $\gamma^{-1} \gamma$ are \emph{not} constant paths, but they are
thinly homotopic to constant paths.  But henceforth, we will relax and
write simply $\gamma$ for the morphism in the path groupoid
corresponding to a path $\gamma$, instead of $[\gamma]$.

As the name suggests, groupoids are a bit like groups. Indeed, a
\define{group} is secretly the same as a groupoid with one object!
In other words, suppose we have group $G$.  Then there is a category
where:
\begin{itemize}
\item There is only one object, $\bullet$.
\item Morphisms from $\bullet$ to $\bullet$ are elements of $G$.
\item Composition of morphisms is multiplication in the group $G$.
\item The identity morphism $1_\bullet$ is the identity element
of $G$.
\end{itemize}
This category is a groupoid, since every group element has an inverse.
Conversely, any groupoid with one object gives a group.  Henceforth
we will freely switch back and forth between thinking of a group 
in the traditional way, and thinking of it as a one-object groupoid.

How can we use groupoids to describe connections? It should not be
surprising that we can do this, now that we have our path groupoid
$\P_1(M)$ and our one-object groupoid $G$ in hand.  A connection
gives a map from $\P_1(M)$ to $G$, which says how to transform 
a particle when we move it along a path.  More precisely: if $G$ is
a Lie group, any connection on the trivial $G$-bundle over $M$
yields a map, called the parallel transport map or
\define{holonomy}, that assigns an element of $G$
to each path: 
\[ \hol \maps \xymatrix{ \bullet 
 & \bullet \ar@/_0.5pc/[l]_{\gamma}  
} 
\quad \mapsto \quad \hol(\gamma) \in G \] 
In physics notation, the holonomy is defined as the 
path-ordered exponential of some $\g$-valued 1-form $A$,
where $\g$ is the Lie algebra of $G$:
\[ \hol(\gamma) = \poexp \left( \int_\gamma A \right) \in G. \]

The holonomy map satisfies certain rules, most of which are summarized
in the word `functor'.  What is a functor?  It is a map between
categories that preserves all the structure in sight!

More precisely: given categories $C$ and $D$, a \define{functor} $F
\maps C \to D$ consists of:
\begin{itemize} 
\item a map $F$ sending objects in $C$ to objects in $D$, and
\item another map, also called $F$, sending morphisms in $C$ to 
morphisms in $D$,
\end{itemize}
such that:
\begin{itemize}
\item given a morphism $f \maps x \to y$ in $C$, we have
$F(f) \maps F(x) \to F(y)$, 
\item $F$ preserves composition:
\[ F(fg) = F(f)F(g) \]
when either side is well-defined, and
\item $F$ preserves identities:
\[ F(1_x) = 1_{F(x)} \]
for every object $x$ of $C$.
\end{itemize}
The last property actually follows from the rest.  The second to 
last---preserving composition---is the most important property
of functors.  As a test of your understanding, check that if $C$ and
$D$ are just \emph{groups} (that is, one-object groupoids) then
a functor $F \maps C \to D$ is just a \emph{homomorphism}.

Let us see what this definition says about a functor
\[       \hol \maps \P_1(M) \to G \]
where $G$ is some Lie group.
This functor $\hol$ must send all the points of $M$ to
the one object of $G$.  More interestingly, it must send
thin homotopy classes of paths in $M$ to elements of $G$:
\[ \hol \maps \xymatrix{ \bullet 
 & \bullet \ar@/_0.5pc/[l]_{\gamma}  
} 
\quad \mapsto \quad \hol(\gamma) \in G \] 
It must preserve composition:
\[ \hol(\gamma \delta) = \hol(\gamma) \, \hol(\delta) \]
and identities:
\[ \hol(1_x) = 1 \in G. \]

While they may be stated in unfamiliar language, these are actually
well-known properties of connections!  First, the holonomy of a
connection along a path
\[ \hol(\gamma) = \poexp \left( \int_\gamma A \right) \in G \]
only depends on the thin homotopy class of $\gamma$.  To 
see this, compute the variation of $\hol(\gamma)$ as we vary the path
$\gamma$, and show the variation is zero if the homotopy is thin.
Second, to compute the group element for a composite of paths, we just
multiply the group elements for each one:
\[ \poexp \left( \int_{\gamma \delta} A \right) = 
\poexp \left( \int_\gamma A \right) \poexp \left( \int_\delta A \right) \]
And third, the path-ordered exponential along a constant path is just the
identity:
\[ \poexp \left( \int_{1_x} A \right) = 1 \in G. \]

All this information is neatly captured by saying $\hol$ is a
functor. And Theorem~\ref{thm:connections} says this is almost all
there is to being a connection. The only additional condition required
is that $\hol$ be \emph{smooth}. This means, roughly, that
$\hol(\gamma)$ depends smoothly on the path $\gamma$---more on that
later.  But if we drop this condition, we can generalize the concept
of connection, and define a \define{generalized connection} on a smooth
manifold $M$ to be a functor $\hol \maps \P_1(M) \to G$. 

Generalized connections have long played an important role in loop
quantum gravity, first in the context of real-analytic manifolds
\cite{AshtekarLewandowski:1994}, and later for smooth manifolds
\cite{BaezSawin:1995,LewandowskiThiemann:1999}.  The reason is that if
$M$ is any manifold and $G$ is a connected compact Lie group, there is
a natural measure on the space of generalized connections.  This means
that you can define a Hilbert space of complex-valued
square-integrable functions on the space of generalized connections.
In loop quantum gravity these are used to describe quantum states
before any constraints have been imposed.  The switch from connections
to generalized connections is crucial here---and the lack of
smoothness gives loop quantum gravity its `discrete' flavor.

But suppose we are interested in ordinary connections.  Then we
really want $\hol(\gamma)$ to depend smoothly on the path $\gamma$.
How can we make this precise?  

One way is to use the theory of `smooth groupoids'
\cite{BaezSchreiber}.  Any Lie group is a smooth groupoid, and so is
the path groupoid of any smooth manifold.  We can define smooth
functors between smooth groupoids, and then smooth functors $\hol
\maps \P_1(M) \to G$ are in one-to-one correspondence with connections
on the trivial principal $G$-bundle over $M$.  We can go even further:
there are more general maps between smooth groupoids, and maps $\hol
\maps \P_1(M) \to G$ of this more general sort correspond to
connections on \textit{not necessarily trivial} principal $G$-bundles
over $M$.  For details, see the work of Bartels \cite{Bartels:2004},
Schreiber and Waldorf \cite{SchreiberWaldorf:functors}.

But if this sounds like too much work, we can take the following
shortcut.  Suppose we have a smooth function $F \maps [0,1]^n \times
[0,1] \to M$, which we think of as a parametrized family of paths.
And suppose that for each fixed value of the parameter $s \in
[0,1]^n$, the path $\gamma_s$ given by
\[               \gamma_s(t) = F(s,t)  \]
is lazy.  Then our functor $\hol \maps \P_1(M) \to G$ gives a function
\[              
\begin{array}{ccl}
[0,1]^n &\to& G  \\
s       &\mapsto& \hol(\gamma_s)  .
\end{array}
\]
If this function is smooth whenever $F$ has the above properties,
then the functor $\hol \maps \P_1(M) \to G$ is {\bf smooth}.

Starting from this definition one can prove the following lemma, 
which lies at the heart of Theorem~\ref{thm:connections}:

\begin{lem}
There is a one-to-one correspondence between smooth functors  
\hfill \break
$\hol \maps \P_1(M) \to G$  and $\Lie(G)$-valued 1-forms $A$ on $M$.
\end{lem}

The idea is that given a $\Lie(G)$-valued 1-form $A$ on $M$, we can 
define a holonomy for any smooth path as follows:
\[ \hol(\gamma) = \poexp\left( \int_\gamma A \right) , \]
and then check that this defines a smooth functor $\hol \maps \P_1(M)
\to G$.  Conversely, suppose we have a smooth functor $\hol$ of this
sort.  Then we can define $\hol(\gamma)$ for smooth paths $\gamma$
that are not lazy, using the fact that every smooth path is thinly
homotopic to a lazy one.  We can even do this for paths $\gamma \maps
[0,s] \to M$ where $s \ne 1$, since any such path can be
reparametrized to give a path of the usual sort.  Given a smooth path
\[ \gamma \maps [0,1] \to M \]
we can truncate it to
obtain a path $\gamma_s$ that goes along $\gamma$ until time $s$:
\[ \gamma_s \maps [0,s] \to M .\]
By what we have said, $\hol(\gamma_s)$ is well-defined.  Using the
fact that $\hol \maps \P_1(M) \to G$ is a smooth functor, one can
check that $\hol(\gamma_s)$ varies smoothly with $s$.  So, we can
differentiate it and define a $\Lie(G)$-valued 1-form $A$ as follows:
\[ A(v) = \frac{d}{ds} \hol(\gamma_s) \big|_{s=0} \]
where $v$ is any tangent vector at a point $x \in M$, and 
$\gamma$ is any smooth path with 
\[        \gamma(0) = x, \qquad \gamma'(0) = v  .\]
Of course, we need to check that $A$ is well-defined and smooth.  We
also need to check that if we start with a smooth functor $\hol$,
construct a 1-form $A$ in this way, and then turn $A$ back into a
smooth functor, we wind up back where we started.

\section{2-Categories and 2-Connections}
\label{2-connections}

Now we want to climb up one dimension, and talk about `2-connections'.
A connection tells us how particles transform as they move along
paths.  A 2-connection will also tell us how \emph{strings} transform as 
they sweep out \emph{surfaces}.  To make this idea precise, we need to take
everything we said in the previous section and boost the dimension
by one.  Instead of categories, we need `2-categories'.  Instead of
groups, we need `2-groups'.  Instead of the path groupoid, we need the
`path 2-groupoid'.  And instead of functors, we need `2-functors'.
When we understand all these things, the analogue of Theorem
\ref{thm:connections} will look strikingly similar to the original
version:

\begin{unnumbered-thm}
For any Lie 2-group $\G$ and any smooth manifold $M$,
there is a one-to-one correspondence between:
\begin{enumerate}
\item 2-connections on the trivial principal $\G$-2-bundle over $M$,
\item pairs consisting of 
a smooth $\g$-valued 1-form $A$ and a smooth $\h$-valued 
2-form $B$ on $M$, such that
\[           \dt(B) = dA + A \wedge A \]
where we use $\dt \maps \h \to \g$, the differential of the map $t \maps
H \to G$, to convert $B$ into a $\g$-valued 2-form, and
\item smooth 2-functors
\[ \hol \maps \P_2(M) \to \G \]
where $\P_2(M)$ is the path 2-groupoid of $M$.
\end{enumerate}
\end{unnumbered-thm}

What does this say?  In brief: there is a way to extract from a Lie
2-group $\G$ a pair of Lie groups $G$ and $H$.  Suppose we have a
1-form $A$ taking values in the Lie algebra of $G$, and a 2-form $B$
valued in the Lie algebra of $H$.  Suppose furthermore that these
forms obey the equation above.  Then we can use
them to consistently define parallel transport, or `holonomies', for
paths and surfaces.  They thus define a `2-connection'.

That is the idea.  But to make it precise, we need 2-categories.

\subsection{2-Categories}

Sets have elements.  Categories have elements, usually
called `objects', but also morphisms between these.  
In an `$n$-category', we go further and include 2-morphisms 
between morphisms, 3-morphisms between 2-morphisms,... and so 
on up to the $n$th level.  We are beginning to see $n$-categories 
provide an algebraic language for $n$-dimensional structures 
in physics \cite{BaezLauda:prehistory}.  Higher gauge theory is just 
one place where this is happening. 

Anyone learning $n$-categories needs to start with 
2-categories \cite{Lack}.  A \define{2-category} consists of:
\begin{itemize}
\item a collection of objects,
\item for any pair of objects $x$ and $y$, a set of morphisms
$f \maps x \to y$:
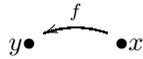
\[
\xymatrix{
   y \bullet 
& \bullet x \ar@/_0.5pc/[l]_{f}
}
\]
\item for any pair of morphisms $f,g \maps x \to y$,
a set of 2-morphisms $\alpha \maps f \To g$:
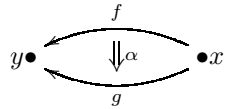
\[
  \xymatrix{
y \bullet 
&& \bullet x
\ar@/^2.5ex/[ll]^{g}="g1"\ar@/_2.5ex/[ll]_{f}="g2"
\ar@{=>}^{\alpha} "g2"+<0ex,-2.5ex>;"g1"+<0ex,2.5ex>
}
\]
We call $f$ the \define{source} of $\alpha$ and $g$ the 
\define{target} of $\alpha$.  
\end{itemize}
Morphisms can be composed just as in a category:
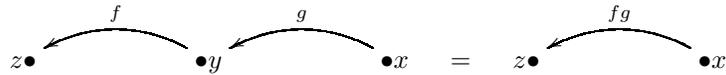
\[
\xymatrix{
   z \bullet 
&& \bullet y \ar@/_1pc/[ll]_{f}
    && \bullet x \ar@/_1pc/[ll]_{g}
}
\quad = \quad
\xymatrix{
   z \bullet 
&& \bullet x \ar@/_1pc/[ll]_{fg}
}
\]
while 2-morphisms can be composed in two distinct ways, vertically:
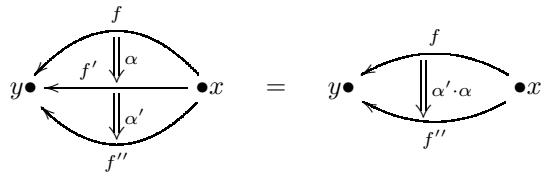
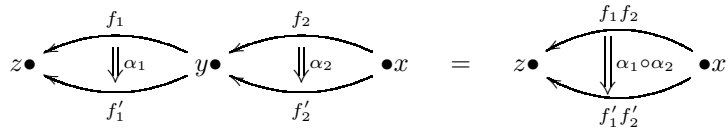
\[
\xymatrix{
   y \bullet && \bullet x
  \ar@/_5ex/[ll]_{f}="g1"
  \ar[ll]_(0.65){f'}
  \ar@{}[ll]|{}="g2"
  \ar@/^5ex/[ll]^{f''}="g3"
  \ar@{=>}^{\alpha} "g1"+<0ex,-2ex>;"g2"+<0ex,0.5ex>
  \ar@{=>}^{\alpha'} "g2"+<0ex,-0.5ex>;"g3"+<0ex,2ex>
}
\quad =\quad
\xymatrix{
y \bullet
&&\bullet x
\ar@/_3ex/[ll]_{f}="g1"
\ar@/^3ex/[ll]^{f''}="g3"
\ar@{=>}^{\alpha'\cdot\alpha} "g1"+<-1ex,-2ex>;"g3"+<-1ex,2ex>
}
\]
and horizontally:
\[
  \xymatrix{
  z \bullet
&&y \bullet
  \ar@/_2.5ex/[ll]_{f_1}="g1"\ar@/^2.5ex/[ll]^{f'_1}="g2"
  \ar@{=>}^{\alpha_1} "g1"+<0ex,-2.5ex>;"g2"+<0ex,2.5ex>
&& \bullet x
  \ar@/_2.5ex/[ll]_{f_2}="g1"\ar@/^2.5ex/[ll]^{f'_2}="g2"
  \ar@{=>}^{\alpha_2} "g1"+<0ex,-2.5ex>;"g2"+<0ex,2.5ex>
}
\quad =\quad
\xymatrix{
 z \bullet
&& \bullet x
\ar@/_3ex/[ll]_{f_1 f_2}="g1"
\ar@/^3ex/[ll]^{f'_1 f_2'}="g3"
\ar@{=>}^{\alpha_1\circ\alpha_2} "g1"+<-1ex,-2ex>;"g3"+<-1ex,2ex>
}
\]
Finally, these laws must hold:
\begin{itemize}
\item Composition of morphisms is associative, and every
object $x$ has a morphism
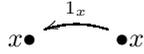
\[
\xymatrix{x \bullet &
\bullet x \ar@/_0.5pc/[l]_{1_x}
}
\]
serving as an identity for composition, just as in an ordinary
category.
\item
Vertical composition is associative, and every morphism $f$ 
has a 2-morphism
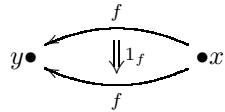
\[
  \xymatrix{
y \bullet 
&& \bullet x
\ar@/^2.5ex/[ll]^{f}="g1"\ar@/_2.5ex/[ll]_{f}="g2"
\ar@{=>}^{1_f} "g2"+<0ex,-2.5ex>;"g1"+<0ex,2.5ex>
}
\]
serving as an identity for vertical composition.
\item
Horizontal composition is associative, and the 2-morphism
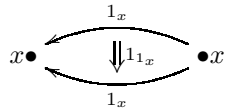
\[
  \xymatrix{
x \bullet 
&& \bullet x
\ar@/^2.5ex/[ll]^{1_x}="g1"\ar@/_2.5ex/[ll]_{1_x}="g2"
\ar@{=>}^{1_{1_x}} "g2"+<0ex,-2.5ex>;"g1"+<0ex,2.5ex>
}
\]
serves as an identity for horizontal composition.
\item
Vertical and horizontal composition of 2-morphisms obey
the \define{interchange law}:
\[
(\alpha'_1 \cdot \alpha_1) \circ (\alpha'_2 \cdot \alpha_2) 
= (\alpha'_1 \circ \alpha'_2) \cdot (\alpha_1 \circ \alpha_2)
\]
so that diagrams of the form
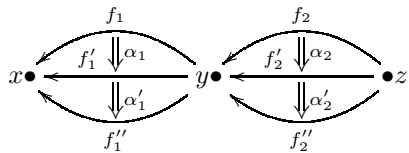
\[
\xymatrix{
  x \bullet 
&& y \bullet
  \ar@/_4ex/[ll]_{f_1}="g1"
  \ar[ll]_(0.65){f'_1}
  \ar@{}[ll]|{}="g2"
  \ar@/^4ex/[ll]^{f''_1}="g3"
  \ar@{=>}^{\alpha_1} "g1"+<0ex,-2ex>;"g2"+<0ex,0.5ex>
  \ar@{=>}^{\alpha'_1} "g2"+<0ex,-0.5ex>;"g3"+<0ex,2ex>
&&  \bullet z
  \ar@/_4ex/[ll]_{f_2}="h1"
  \ar[ll]_(0.65){f'_2}
  \ar@{}[ll]|{}="h2"
  \ar@/^4ex/[ll]^{f''_2}="h3"
  \ar@{=>}^{\alpha_2} "h1"+<0ex,-2ex>;"h2"+<0ex,0.5ex>
  \ar@{=>}^{\alpha'_2} "h2"+<0ex,-0.5ex>;"h3"+<0ex,2ex>
}
\]
define unambiguous 2-morphisms.
\end{itemize}

The interchange law is the truly new thing here.  A category is
all about attaching 1-dimensional arrows end to end, and we need the
associative law to do that unambiguously.  In a 2-category, we
visualize the 2-morphisms as little pieces of 2-dimensional surface:
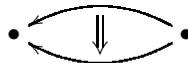
\[
  \xymatrix{
 \bullet 
&& \bullet 
\ar@/^2.5ex/[ll]^{\;}="g1"\ar@/_2.5ex/[ll]_{\;}="g2"
\ar@{=>}^{\;} "g2"+<0ex,-1.5ex>;"g1"+<0ex,1.5ex>
}
\]
We can attach these together in two ways: vertically and
horizontally.  For the result to be unambiguous, we need not only
associative laws but also the interchange law.  In what follows we
will see this law turning up all over the place.

\subsection{Path 2-Groupoids}

Path groupoids play a big though often neglected role in physics: the
path groupoid of a spacetime manifold describes all the possible
motions of a point particle in that spacetime.  The path 2-groupoid
does the same thing for particles \textit{and strings}.

First of all, a {\bf 2-groupoid} is a 2-category where:
\begin{itemize}
\item
Every morphism $f \maps x \to y$ has an \define{inverse},
$f^{-1} \maps y \to x$, such that:
\[      f^{-1} f = 1_x  
\quad \textrm{and} \quad
f f^{-1} = 1_y . \]
\item
Every 2-morphism $\alpha \maps f \To g$ has a \define{vertical inverse},
$\alpha^{-1}_{\textrm{\small vert}} \maps g \To f$, such that:
\[     \alpha^{-1}_{\textrm{\small vert}} \cdot \alpha = 1_f   
\quad \textrm{and} \quad
       \alpha \cdot \alpha^{-1}_{\textrm{\small vert}} = 1_f . \]
\end{itemize}
It actually follows from this definition that every 2-morphism 
$\alpha \maps f \To g$ also has a \define{horizontal inverse}, 
$\alpha^{-1}_{\textrm{\small hor}} \maps f^{-1} \To g^{-1}$, such that:
\[     \alpha^{-1}_{\textrm{\small hor}} \circ \alpha = 1_{1_x}  
\quad \textrm{and} \quad
       \alpha \circ \alpha^{-1}_{\textrm{\small hor}} = 1_{1_x} . \]
So, a 2-groupoid has every kind of inverse your heart could desire.

An example of a 2-group is the `path 2-groupoid' of a smooth manifold
$M$.  To define this, we can start with the path groupoid $\P_1(M)$ as
defined in the previous section, and then throw in 2-morphisms.  Just
as the morphisms in $\P_1(M)$ were thin homotopy classes of lazy paths, 
these 2-morphisms will be thin homotopy classes of lazy surfaces.

What is a `lazy surface'?   First, recall that a \define{homotopy}
between lazy paths $\gamma, \delta \maps x \to y$ is a smooth map
$\Sigma \maps [0,1]^2 \to M$ with
\[   \Sigma(0,t) = \gamma(t)  \]
\[   \Sigma(1,t) = \delta(t)  \] 
We say this homotopy is a \define{lazy surface} if 
\begin{itemize}
\item
$\Sigma(s,t)$ is independent of $s$ near $s = 0$ and near $s = 1$, 
\item
$\Sigma(s,t)$ is constant near $t = 0$ and constant near $t = 1$.
\end{itemize}
Any homotopy $\Sigma$ yields a one-parameter family of paths 
$\gamma_s$ given by
\[             \gamma_s(t) = \Sigma(s,t) .\]
If $\Sigma$ is a lazy surface, each of these paths is lazy.
Furthermore, the path $\gamma_s$ equals $\gamma_0$ when $s$ is
sufficiently close to $0$, and it equals $\gamma_1$ when $s$ is
sufficiently close to $1$.  This allows us to compose lazy homotopies
either vertically or horizontally and obtain new lazy homotopies!

However, vertical and horizontal composition will only obey the
2-groupoid axioms if we take 2-morphisms in the path 2-groupoid to be
\emph{equivalence} classes of lazy surfaces.  We saw this kind of
issue already when discussing the path groupoid, so we we will allow
ourselves to be a bit sketchy this time.  The key idea is to define a
concept of `thin homotopy' between lazy surfaces $\Sigma$ and $\Xi$.
For starters, this should be a smooth map $H \maps [0,1]^3 \to M$ such
that $H(0,s,t) = \Sigma(s,t)$ and $H(1,s,t) = \Xi(s,t)$.  But we also
want $H$ to be `thin'.  In other words, it should sweep out no volume:
the rank of the differential $dH$ should be less than $3$ at every
point.  

To make thin homotopies well-defined between thin homotopy classes of
paths, some more technical conditions are also useful.  For these, the
reader can turn to Section 2.1 of Schreiber and Waldorf
\cite{SchreiberWaldorf:functors}.  The upshot is that we obtain
for any smooth manifold $M$ a \define{path 2-groupoid} $\P_2(M)$, 
in which:
\begin{itemize}
\item An object is a point of $M$.
\item A morphism from $x$ to $y$ is a thin homotopy class of lazy paths
from $x$ to $y$.  
\item A 2-morphism between equivalence classes of lazy paths
$\gamma_0, \gamma_1 \maps x \to y$ is a thin homotopy class of
lazy surfaces $\Sigma \maps \gamma_0 \To \gamma_1$.  
\end{itemize}
As we already did with the concept of `lazy path', we will often use
`lazy surface' to mean a thin homotopy class of lazy surfaces.  But 
now let us hasten on to another important class of 2-groupoids, the 
`2-groups'.  Just as groups describe symmetries in gauge theory, these
describe symmetries in higher gauge theory.

\subsection{2-Groups}

Just as a group was a groupoid with one object, we define a
\define{2-group} to be a 2-groupoid with one object.  This definition
is so elegant that it may be hard to understand at first!  So, it will
be useful to take a 2-group $\G$ and chop it into four bite-sized
pieces of data, giving a `crossed module' $(G,H,t,\alpha)$.  Indeed,
2-groups were originally introduced in the guise of crossed modules by
the famous topologist J.\ H.\ C.\ Whitehead \cite{Whitehead}.  In
1950, with help from Mac Lane \cite{MW}, he used crossed modules to
generalize the fundamental group of a space to what we might now call
the `fundamental 2-group'.  But only later did it become clear that a
crossed module was another way of talking about a 2-groupoid with just
one object!  For more of this history, and much more on 2-groups, see
\cite{BaezLauda:2groups}.

Let us start by seeing what it means to say a 2-group is a
2-groupoid with one object.  It means that a 2-group $\G$ has:
\begin{itemize}
\item one object: 
\[ \bullet \]
\item morphisms: 
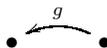
\[
\xymatrix{
 \bullet 
& \bullet \ar@/_0.5pc/[l]_{g}
}
\]
\item and 2-morphisms:
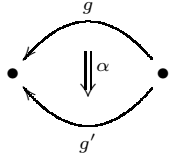
\[
\xy 
(-10,0)*+{\bullet}="4"; 
(10,0)*+{\bullet}="6";
{\ar@/_1.65pc/_{g} "6";"4"}; 
{\ar@/^1.65pc/^{g'} "6";"4"};
{\ar@{=>}^<<<{\alpha} (0,3)*{};(0,-3)*{}} ;
\endxy 
\]
\end{itemize}
The morphisms form a group under composition:
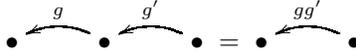
\[ \xymatrix{ \bullet &
\ar@/_/[l]_g  
\bullet 
& 
\bullet 
\ar@/_/[l]_{g'} 
} = 
\xymatrix{ \bullet & \bullet \ar@/_/[l]_{gg'} } \]
The 2-morphisms form a group under horizontal composition:
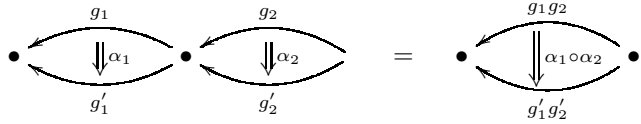
\[
  \xymatrix{
   \bullet
&& \bullet
  \ar@/_2.5ex/[ll]_{g_1}="g1"\ar@/^2.5ex/[ll]^{g'_1}="g2"
  \ar@{=>}^{\alpha_1} "g1"+<0ex,-2.5ex>;"g2"+<0ex,2.5ex>
&& 
  \ar@/_2.5ex/[ll]_{g_2}="g1"\ar@/^2.5ex/[ll]^{g'_2}="g2"
  \ar@{=>}^{\alpha_2} "g1"+<0ex,-2.5ex>;"g2"+<0ex,2.5ex>
}
\quad =\quad
\xymatrix{
  \bullet
&& \bullet 
\ar@/_3ex/[ll]_{g_1 g_2}="g1"
\ar@/^3ex/[ll]^{g'_1 g_2'}="g3"
\ar@{=>}^{\alpha_1\circ\alpha_2} "g1"+<-1ex,-2ex>;"g3"+<-1ex,2ex>
}
\]
In addition, the 2-morphisms can be composed vertically:
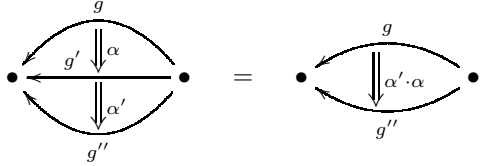
\[
\xymatrix{
    \bullet && \bullet 
  \ar@/_5ex/[ll]_{g}="g1"
  \ar[ll]_(0.65){g'}
  \ar@{}[ll]|{}="g2"
  \ar@/^5ex/[ll]^{g''}="g3"
  \ar@{=>}^{\alpha} "g1"+<0ex,-2ex>;"g2"+<0ex,0.5ex>
  \ar@{=>}^{\alpha'} "g2"+<0ex,-0.5ex>;"g3"+<0ex,2ex>
}
\quad =\quad
\xymatrix{
\bullet
&&\bullet 
\ar@/_3ex/[ll]_{g}="g1"
\ar@/^3ex/[ll]^{g''}="g3"
\ar@{=>}^{\alpha'\cdot\alpha} "g1"+<-1ex,-2ex>;"g3"+<-1ex,2ex>
}
\]
Vertical composition is also associative with identity and
inverses.  But the 2-morphisms do not form a group under this operation,
because a given pair may not be composable: their source
and target may not match up.  Finally, vertical and horizontal 
composition are tied together by the interchange law, which says
the two ways one can read this diagram are consistent.
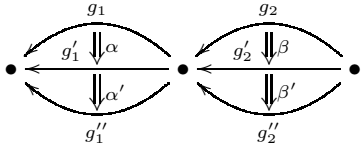
\[
\xymatrix{
   \bullet 
&&  \bullet
  \ar@/_4ex/[ll]_{g_1}="g1"
  \ar[ll]_(0.65){g'_1}
  \ar@{}[ll]|{}="g2"
  \ar@/^4ex/[ll]^{g''_1}="g3"
  \ar@{=>}^{\alpha} "g1"+<0ex,-2ex>;"g2"+<0ex,0.5ex>
  \ar@{=>}^{\alpha'} "g2"+<0ex,-0.5ex>;"g3"+<0ex,2ex>
&&  \bullet 
  \ar@/_4ex/[ll]_{g_2}="h1"
  \ar[ll]_(0.65){g'_2}
  \ar@{}[ll]|{}="h2"
  \ar@/^4ex/[ll]^{g''_2}="h3"
  \ar@{=>}^{\beta} "h1"+<0ex,-2ex>;"h2"+<0ex,0.5ex>
  \ar@{=>}^{\beta'} "h2"+<0ex,-0.5ex>;"h3"+<0ex,2ex>
}
\]

Now let us create a crossed module $(G,H,t,\alpha)$ from a 2-group $\G$.  
To do this, first note that the morphisms of the 2-group form a group by 
themselves, with composition as the group operation.  So:
\begin{itemize}
\item Let $G$ be the set of morphisms in $\G$, made into a
group with composition as the group operation:
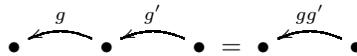
\[ \xymatrix{ \bullet &
\ar@/_/[l]_g  
\bullet 
& 
\bullet 
\ar@/_/[l]_{g'} 
} = 
\xymatrix{ \bullet & \bullet \ar@/_/[l]_{gg'} } \]
\end{itemize}
How about the 2-morphisms?  These also form a group, with horizontal 
composition as the group operation.  But it turns out to be 
efficient to focus on a subgroup of this:
\begin{itemize}
\item Let $H$ be the set of all 2-morphisms 
whose source is the identity:
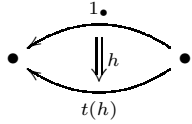
\[
\xymatrix{\bullet
&& \bullet
\ar@/_3ex/[ll]_{1_\bullet}="g1"
\ar@/^3ex/[ll]^{t(h)}="g3"
\ar@{=>}^{h} "g1"+<0ex,-2.5ex>;"g3"+<0ex,2.5ex>
}
\]
We make $H$ 
into a group with horizontal composition as the group 
operation:
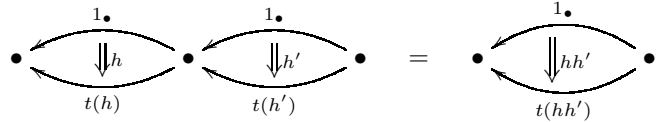
\[
  \xymatrix{
  \bullet
&&\bullet
  \ar@/_2.5ex/[ll]_{1_\bullet}="g1"\ar@/^2.5ex/[ll]^{t(h)}="g2"
  \ar@{=>}^{h} "g1"+<0ex,-2.5ex>;"g2"+<0ex,2.5ex>
&&\bullet
  \ar@/_2.5ex/[ll]_{1_\bullet}="g1"\ar@/^2.5ex/[ll]^{t(h')}="g2"
  \ar@{=>}^{h'} "g1"+<0ex,-2.5ex>;"g2"+<0ex,2.5ex>
}
\quad =\quad
\xymatrix{
\bullet
&& \bullet
\ar@/_3ex/[ll]_{1_\bullet}="g1"
\ar@/^3ex/[ll]^{t(hh')}="g3"
\ar@{=>}^{hh'} "g1"+<-1ex,-2.5ex>;"g3"+<-1ex,2.5ex>
}
\]
\end{itemize}
Above we use $hh'$ as an abbreviation for the horizontal composite $h
\circ h'$ of two elements of $H$.  We will use $h^{-1}$ to denote the
horizontal inverse of an element of $H$.  We use $t(h)$ to denote the
target of an element $h \in H$.  The definition of a 2-category
implies that $t \maps H \to G$ is a group homomorphism:
\[            t(hh') = t(h) t(h')  .\]
This homomorphism is our third piece of data:
\begin{itemize}
\item
A group homomorphism $t \maps H \to G$ sending each 2-morphism in
$H$ to its target:
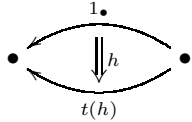
\[
\xymatrix{\bullet
&& \bullet
\ar@/_3ex/[ll]_{1_\bullet}="g1"
\ar@/^3ex/[ll]^{t(h)}="g3"
\ar@{=>}^{h} "g1"+<0ex,-2.5ex>;"g3"+<0ex,2.5ex>
}
\]
\end{itemize}
The fourth piece of data is the subtlest.  There is a way
to `horizontally conjugate' any element $h \in H$ by an element 
$g \in G$, or more precisely by its identity 2-morphism $1_g$:
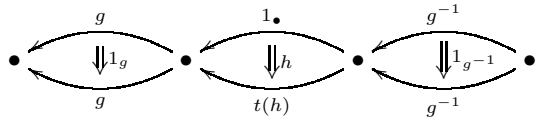
\[
  \xymatrix{
  \bullet
&&\bullet
  \ar@/_2.5ex/[ll]_{g}="g1"\ar@/^2.5ex/[ll]^{g}="g2"
  \ar@{=>}^{1_g} "g1"+<0ex,-2.5ex>;"g2"+<0ex,2.5ex>
&&\bullet
  \ar@/_2.5ex/[ll]_{1_\bullet}="g1"\ar@/^2.5ex/[ll]^{t(h)}="g2"
  \ar@{=>}^{h} "g1"+<0ex,-2.5ex>;"g2"+<0ex,2.5ex>
&&\bullet
  \ar@/_2.5ex/[ll]_{g^{-1}}="g1"\ar@/^2.5ex/[ll]^{g^{-1}}="g2"
  \ar@{=>}^{1_{g^{-1}}} "g1"+<0ex,-2.5ex>;"g2"+<0ex,2.5ex>
}
\]
The result is a 2-morphism in $H$ which we call $\alpha(g)(h)$.  In
fact, $\alpha(g)$ is an {\bf automorphism} of
$H$, meaning a one-to-one and onto function with
\[             \alpha(g)(hh') = \alpha(g)(h) \; \alpha(g)(h')  .\]
Composing two automorphisms gives another automorphism, and this
makes the automorphisms of $H$ into a group, say $\Aut(H)$.  Even
better, $\alpha$ gives a group homomorphism
\[        \alpha \maps G \to \Aut(H) . \]
Concretely, this means that in addition to the above equation, we have
\[        \alpha(g g') = \alpha(g)\alpha(g'). \]
Checking these two equations is a nice way to test your
understanding of 2-categories.  A group homomorphism $\alpha
\maps G \to \Aut(H)$ is also called an \define{action} of the group
$G$ on the group $H$.  So, the fourth and final piece of data in our
crossed module is:
\begin{itemize}
\item
An action $\alpha$ of $G$ on $H$ given by:
\[
  \xymatrix{
  \bullet
&&\bullet
  \ar@/_2.5ex/[ll]_{g}="g1"\ar@/^2.5ex/[ll]^{g}="g2"
  \ar@{=>}^{1_g} "g1"+<0ex,-2.5ex>;"g2"+<0ex,2.5ex>
&&\bullet
  \ar@/_2.5ex/[ll]_{1_\bullet}="g1"\ar@/^2.5ex/[ll]^{t(h)}="g2"
  \ar@{=>}^{h} "g1"+<0ex,-2.5ex>;"g2"+<0ex,2.5ex>
&&\bullet
  \ar@/_2.5ex/[ll]_{g^{-1}}="g1"\ar@/^2.5ex/[ll]^{g^{-1}}="g2"
  \ar@{=>}^{1_g^{-1}} "g1"+<0ex,-2.5ex>;"g2"+<0ex,2.5ex>
}
\quad = \quad
  \xymatrix{
  \bullet
&&\bullet
  \ar@/_2.5ex/[ll]_{1}="g1"\ar@/^2.5ex/[ll]^{t(\alpha(g)(h))}="g2"
  \ar@{=>}^{\alpha(g)(h)} "g1"+<-1ex,-2.5ex>;"g2"+<-1ex,2.5ex>
}
\]
\end{itemize}
A crossed module $(G,H,t,\alpha)$ must also satisfy two more
equations which follow from the definition of a 2-group.
First, examining the above diagram, we see that $t$ is
\define{\boldmath$G$-equivariant}, by which we mean:
\begin{itemize}
\item $t(\alpha(g)h) = g ( t(h) ) g^{-1}$ for all $g \in G$
and $h \in H$.
\end{itemize}
Second, the \define{Peiffer identity} holds:
\begin{itemize}
\item $ \alpha(t(h))h' = hh'h^{-1}$ for all $h,h' \in H$.
\end{itemize}

The Peiffer identity is the least obvious thing about a crossed
module.  It follows from the interchange law, and it is worth
seeing how.  First, we have:
\[
hh'h^{-1} \quad = \quad
  \xymatrix{
  \bullet
&&\bullet
  \ar@/_3ex/[ll]_{1_\bullet}="g1"\ar@/^3ex/[ll]^{t(h)}="g2"
  \ar@{=>}^{h} "g1"+<0ex,-2.5ex>;"g2"+<0ex,2.5ex>
&&\bullet
  \ar@/_3ex/[ll]_{1_\bullet}="g1"\ar@/^3ex/[ll]^{t(h')}="g2"
  \ar@{=>}^{h'} "g1"+<0ex,-2.5ex>;"g2"+<0ex,2.5ex>
&&\bullet
  \ar@/_3ex/[ll]_{1_\bullet}="g1"\ar@/^3ex/[ll]^{t(h^{-1})}="g2"
  \ar@{=>}^{h^{-1}} "g1"+<0ex,-2.5ex>;"g2"+<0ex,2.5ex>
}
\]
where---beware!---we are now using $h^{-1}$ to mean the
\emph{horizontal} inverse of $h$, since this is its inverse in the
group $H$.  We can pad out this equation by vertically composing with
some identity morphisms:
\[
hh'h^{-1} \quad = \quad
\xymatrix{
   \bullet 
&&&  \bullet
  \ar@/_6ex/[lll]_{1_\bullet}="g1"
  \ar[lll]_(0.7){t(h)}
  \ar@{}[lll]|{}="g2"
  \ar@/^6ex/[lll]^{t(h)}="g3"
  \ar@{=>}^{h} "g1"+<0ex,-2ex>;"g2"+<0ex,0.5ex>
  \ar@{=>}^{1_{t(h)}} "g2"+<0ex,-0.5ex>;"g3"+<0ex,2ex>
&&&  \bullet 
  \ar@/_6ex/[lll]_{1_\bullet}="h1"
  \ar[lll]_(0.7){1_\bullet}
  \ar@{}[lll]|{}="h2"
  \ar@/^6ex/[lll]^{t(h)}="h3"
  \ar@{=>}^{1_{1_\bullet}} "h1"+<0ex,-2ex>;"h2"+<0ex,0.5ex>
  \ar@{=>}^{h} "h2"+<0ex,-0.5ex>;"h3"+<0ex,2ex>
&&&  \bullet
  \ar@/_6ex/[lll]_{1_\bullet}="g1"
  \ar[lll]_(0.70){t(h^{-1})}
  \ar@{}[lll]|{}="g2"
  \ar@/^6ex/[lll]^{t(h^{-1})}="g3"
  \ar@{=>}^{h^{-1}} "g1"+<-0.5ex,-2ex>;"g2"+<-0.5ex,0.5ex>
  \ar@{=>}^{1_{t(h^{-1})}} "g2"+<-0.5ex,-0.5ex>;"g3"+<-0.5ex,2ex>
}
\]
This diagram describes an unambiguous 2-morphism,
thanks to the interchange law.  So, we can do the horizontal
compositions first and get:
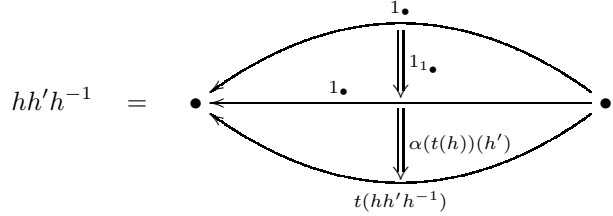
\[
hh'h^{-1} \quad = \quad
\xymatrix{
   \bullet 
&&&&&  \bullet
  \ar@/_7ex/[lllll]_{1_\bullet}="g1"
  \ar[lllll]_(0.65){1_\bullet}
  \ar@{}[lllll]|{}="g2"
  \ar@/^7ex/[lllll]^{t(hh'h^{-1})}="g3"
  \ar@{=>}^{1_{1_\bullet}} "g1"+<0ex,-2ex>;"g2"+<0ex,0.5ex>
  \ar@{=>}^{\alpha(t(h))(h')} "g2"+<0ex,-0.5ex>;"g3"+<0ex,2ex>
}
\]
But vertically composing with an identity 2-morphism has no
effect.  So, we obtain the Peiffer identity:
\[  hh'h^{-1} = \alpha(t(h))(h')  . \]

All this leads us to define a \define{crossed module} $(G,H,t,\alpha)$ 
to consist of:
\begin{itemize}
\item a group $G$,
\item a group $H$, 
\item a homomorphism $t \maps H \to G$, and
\item an action $\alpha \maps G \to \Aut(H)$ 
\end{itemize}
such that:
\begin{itemize} 
\item $t$ is $G$-equivariant: 
\[  t(\alpha(g)h) = g ( t(h) ) g^{-1}  \]
for all $g \in G$ and $h \in H$, and
\item
the Peiffer identity holds for all $h, h'\in H$:
\[  \alpha(t(h))h' = hh'h^{-1} .\]
\end{itemize}

In fact, we can recover a 2-group $\G$ from its crossed module
$(G,H,t,\alpha)$, so crossed modules are just another way of thinking
about 2-groups.  The trick to seeing this is to notice that 
2-morphisms in $\G$ are the same as pairs $(g,h) \in G \times H$.
Such a pair gives this 2-morphism:
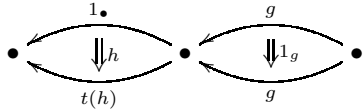
\[          
  \xymatrix{
   \bullet
&& \bullet
  \ar@/_2.5ex/[ll]_{1_\bullet}="g1"\ar@/^2.5ex/[ll]^{t(h)}="g2"
  \ar@{=>}^{h} "g1"+<0ex,-2.5ex>;"g2"+<0ex,2.5ex>
&& \bullet
  \ar@/_2.5ex/[ll]_{g}="g1"\ar@/^2.5ex/[ll]^{g}="g2"
  \ar@{=>}^{1_g} "g1"+<0ex,-2.5ex>;"g2"+<0ex,2.5ex>
}
\]
We leave it to the reader to check that every 2-morphism in $\G$ is of
this form.  Note that this 2-morphism goes from $g$ to $t(h)g$.  So,
when we construct a 2-group from a crossed module, we get a
2-morphism 
\[             (g,h) \maps g \to t(h) g  \]
from any pair $(g,h) \in G \times H$.  Horizontal composition of
2-morphisms then makes $G \times H$ into a group, as follows:
\[ (g,h) \circ (g',h') = \]
\[
  \xymatrix{
   \bullet
& \bullet
  \ar@/_2.5ex/[l]_{1_\bullet}="g1"\ar@/^2.5ex/[l]^{t(h)}="g2"
  \ar@{=>}^{h} "g1"+<0ex,-2.5ex>;"g2"+<0ex,2.5ex>
& \bullet
  \ar@/_2.5ex/[l]_{g}="g1"\ar@/^2.5ex/[l]^{g}="g2"
  \ar@{=>}^{1_g} "g1"+<0ex,-2.5ex>;"g2"+<0ex,2.5ex>
& \bullet
  \ar@/_2.5ex/[l]_{1_\bullet}="g1"\ar@/^2.5ex/[l]^{t(h')}="g2"
  \ar@{=>}^{h'} "g1"+<0ex,-2.5ex>;"g2"+<0ex,2.5ex>
& \bullet
  \ar@/_2.5ex/[l]_{g'}="g1"\ar@/^2.5ex/[l]^{g'}="g2"
  \ar@{=>}^{1_{g'}} "g1"+<0ex,-2.5ex>;"g2"+<0ex,2.5ex>
}
= \]
\[
  \xymatrix{
   \bullet
& \bullet
  \ar@/_2.5ex/[l]_{1_\bullet}="g1"\ar@/^2.5ex/[l]^{t(h)}="g2"
  \ar@{=>}^{h} "g1"+<0ex,-2.5ex>;"g2"+<0ex,2.5ex>
& \bullet
  \ar@/_2.5ex/[l]_{g}="g1"\ar@/^2.5ex/[l]^{g}="g2"
  \ar@{=>}^{1_g} "g1"+<0ex,-2.5ex>;"g2"+<0ex,2.5ex>
& \bullet
  \ar@/_2.5ex/[l]_{1_\bullet}="g1"\ar@/^2.5ex/[l]^{t(h')}="g2"
  \ar@{=>}^{h'} "g1"+<0ex,-2.5ex>;"g2"+<0ex,2.5ex>
& \bullet
  \ar@/_2.5ex/[l]_{g^{-1}}="g1"\ar@/^2.5ex/[l]^{g^{-1}}="g2"
  \ar@{=>}^{1_{g^{-1}}} "g1"+<-1ex,-2.5ex>;"g2"+<-1ex,2.5ex>
& \bullet
  \ar@/_2.5ex/[l]_{g}="g1"\ar@/^2.5ex/[l]^{g}="g2"
  \ar@{=>}^{1_{g}} "g1"+<0ex,-2.5ex>;"g2"+<0ex,2.5ex>
& \bullet
  \ar@/_2.5ex/[l]_{g'}="g1"\ar@/^2.5ex/[l]^{g'}="g2"
  \ar@{=>}^{1_{g'}} "g1"+<0ex,-2.5ex>;"g2"+<0ex,2.5ex>
}
= 
\]
\[
   \xymatrix{
   \bullet
& \bullet
  \ar@/_2.5ex/[l]_{1_\bullet}="g1"\ar@/^2.5ex/[l]^{t(h)}="g2"
  \ar@{=>}^{h} "g1"+<0ex,-2.5ex>;"g2"+<0ex,2.5ex>
&&& \bullet
  \ar@/_2.5ex/[lll]_{1_\bullet}="g1"\ar@/^2.5ex/[lll]^{t(\alpha(g)(h'))}="g2"
  \ar@{=>}^{\alpha(g)(h')} "g1"+<0ex,-2.5ex>;"g2"+<0ex,2.5ex>
&& \bullet
  \ar@/_2.5ex/[ll]_{gg'}="g1"\ar@/^2.5ex/[ll]^{gg'}="g2"
  \ar@{=>}^{1_{gg'}} "g1"+<0ex,-2.5ex>;"g2"+<0ex,2.5ex>
}
= \]
\[
  \xymatrix{
   \bullet
&&&& \bullet
  \ar@/_4ex/[llll]_{1_\bullet}="g1"\ar@/^4ex/[llll]^{t(h\alpha(g)(h'))}="g2"
  \ar@{=>}^{h\alpha(g)(h')} "g1"+<0ex,-2.5ex>;"g2"+<0ex,2.5ex>
&& \bullet
  \ar@/_2.5ex/[ll]_{gg'}="g1"\ar@/^2.5ex/[ll]^{gg'}="g2"
  \ar@{=>}^{1_{gg'}} "g1"+<0ex,-2.5ex>;"g2"+<0ex,2.5ex>
} 
= 
(gg', h\alpha(g)(h')) .\]
So, the group of 2-morphisms of $\G$ is the semidirect product 
$G \ltimes H$, defined using the action $\alpha$.

Following this line of thought, the reader can check the following:

\begin{thm}
\label{crossed_module}
Given a crossed module $(G,H,t,\alpha)$, there is a unique 2-group
$\G$ where:
\begin{itemize}
\item
the group of morphisms is $G$,
\item
a 2-morphism $\alpha \maps g \To g'$ is the same
as a pair $(g,h) \in G \times H$ with $g' = t(h)g$,
\item
the vertical composite of $(g,h)$ and $(g',h')$, when they
are composable, is given by
\[   (g,h) \cdot (g',h') = (g',hh') , \]
\item 
the horizontal composite of $(g,h)$ and $(g',h')$ is given by
\[   (g,h) \circ (g',h') = (gg', h \alpha(g)(h')) .\]
\end{itemize}
Conversely, given a 2-group $\G$, there is a unique crossed module
$(G,H,t,\alpha)$ where:
\begin{itemize}
\item $G$ is the group of morphisms of $\G$, 
\item $H$ is the group of 2-morphisms with source equal to $1_\bullet$,
\item $t \maps H \to G$ assigns to each 2-morphism in $H$ its target,
\item the action $\alpha$ of $G$ on $H$ is given by
\[           \alpha(g) h = 1_g \circ h \circ 1_{g^{-1}}\;  .\]
\end{itemize}
\end{thm}

Indeed, these two processes set up an equivalence between 2-groups and
crossed modules, as described more formally elsewhere
\cite{BaezLauda:2groups,ForresterBarker}.  It thus makes sense to
define a \define{Lie 2-group} to be a 2-group for which the groups $G$
and $H$ in its crossed module are Lie groups, with the maps $t \maps H
\to G$ and $\alpha \maps G \to \Aut(H)$ being smooth.  It is worth
emphasizing that in this context we use $\Aut(H)$ to mean the group of
\emph{smooth} automorphisms of $H$.  This is a Lie group in its own
right.

In Section \ref{examples} we will use Theorem \ref{crossed_module} to
construct many examples of Lie 2-groups.  But first we should finish
explaining 2-connections. 

\subsection{2-Connections}

A 2-connection is a recipe for parallel transporting both
0-dimensional and 1-dimensional objects---say, particles and
strings.  Just as we can describe a connection on a trivial bundle
using a Lie-algebra valued differential form, we can describe a
2-connection using a \emph{pair} of differential forms.  But there is
a deeper way of understanding 2-connections.  Just as a connection was
revealed to be a smooth functor
\[           \hol \maps \P_1(M) \to G  \]
for some Lie group $G$, a 2-connection will turn out to be a 
smooth 2-functor 
\[   \hol \maps \P_2(M) \to \G \]
for some Lie 2-group $\G$.  Of course, to make sense of this we need
to define a `2-functor', and say what it means for such a thing to be
smooth.

The definition of 2-functor is utterly straightforward: it is 
a map between 2-categories that preserves everything in sight.
So, given 2-categories $C$ and $D$, a \define{2-functor} 
$F \maps C \to D$ consists of:
\begin{itemize} 
\item a map $F$ sending objects in $C$ to objects in $D$, 
\item another map called $F$ sending morphisms in $C$ to 
morphisms in $D$,
\item a third map called $F$ sending 2-morphisms in $C$ to 
2-morphisms in $D$,
\end{itemize}
such that:
\begin{itemize}
\item given a morphism $f \maps x \to y$ in $C$, we have
$F(f) \maps F(x) \to F(y)$, 
\item $F$ preserves composition for morphisms, and identity morphisms:
\[ F(fg) = F(f)F(g) \]
\[ F(1_x) = 1_{F(x)} ,\]
\item given a 2-morphism $\alpha \maps f \To g$ in $C$, we have
$F(\alpha) \maps F(f) \To F(g)$, 
\item $F$ preserves vertical and horizontal
composition for 2-morphisms, and identity 2-morphisms:
\[ F(\alpha \cdot \beta ) = F(\alpha) \cdot F(\beta) \]
\[ F(\alpha \circ \beta ) = F(\alpha) \circ F(\beta) \]
\[ F(1_f) = 1_{F(f)}. \]
\end{itemize}

There is a general theory of smooth 2-groupoids and smooth 2-functors
\cite{BaezSchreiber, SchreiberWaldorf:gerbes}.  But here we prefer to take a more
elementary approach.  We already know that for any Lie 2-group $\G$, the morphisms
and 2-morphisms each form Lie groups. Given this, we can say that for any smooth
manifold $M$, a 2-functor
\[   \hol \maps \P_2(M) \to \G \]
is \define{smooth} if:
\begin{itemize}
\item For any smoothly parametrized family of lazy paths $\gamma_s$
($s \in [0,1]^n$) the morphism $\hol(\gamma_s)$ depends smoothly
on $s$, and
\item For any smoothly parametrized family of lazy surfaces $\Sigma_s$
($s \in [0,1]^n$) the 2-morphism $\hol(\Sigma_s)$ depends smoothly
on $s$.
\end{itemize}

With these definitions in hand, we are finally ready to understand
the basic result about 2-connections.  It is completely analogous to
Theorem~\ref{thm:connections}:

\begin{thm}
\label{thm:2-connections}
For any Lie 2-group $\G$ and any smooth manifold $M$,
there is a one-to-one correspondence between:
\begin{enumerate}
\item 2-connections on the trivial principal $\G$-2-bundle over $M$,
\item pairs consisting of 
a smooth $\g$-valued 1-form $A$ and a smooth $\h$-valued 
2-form $B$ on $M$, such that
\[           \dt(B) = dA + A \wedge A \]
where we use $\dt \maps \h \to \g$, the differential of the map $t \maps
H \to G$, to convert $B$ into a $\g$-valued 2-form, and
\item smooth 2-functors
\[ \hol \maps \P_2(M) \to \G \]
where $\P_2(M)$ is the path 2-groupoid of $M$.
\end{enumerate}
\end{thm}

This result was announced by Baez and Schreiber \cite{BaezSchreiber},
and a proof can be be found in the work of Schreiber and Waldorf
\cite{SchreiberWaldorf:gerbes}.  This work was deeply inspired by
the ideas of Breen and Messing \cite{Breen:2008, BreenMessing}, who
considered a special class of 2-groups, and omitted the equation
$\dt(B) = dA + A \wedge A$, since their sort of connection did not assign
holonomies to surfaces.  One should also compare the closely 
related work of Mackaay, Martins, and Picken \cite{MackaayPicken, 
MartinsPicken:2007}, and the work of Pfeiffer and Girelli 
\cite{Pfeiffer:2003, GirelliPfeiffer:2004}.

In the above theorem, the first item mentions `2-connections' and
`2-bundles'---concepts that we have not defined.  But since we are
only talking about 2-connections on \emph{trivial} 2-bundles, we do
not need these general concepts yet.  For now, we can take the third
item as the definition of the first.  Then the content of the theorem
lies in the differential form description of smooth 2-functors $\hol
\maps \P_2(M) \to \G$.  This is what we need to understand.

A 2-functor of this sort must assign holonomies both to paths and
surfaces.  As you might expect, the 1-form $A$ is primarily
responsible for defining holonomies along paths, while the 2-form $B$
is responsible for defining holonomies for surfaces.  But this is a
bit of an oversimplification.  When computing the holonomy of a surface,
we need to use $A$ as well as $B$! 

Another surprising thing is that $A$ and $B$ need to be related by an
\textit{equation} for the holonomy to be a 2-functor.  If we ponder
how the holonomy of a surface is actually computed, we can see why
this is so.  We shall not be at all rigorous here.  We just want
to give a rough intuitive idea of how to compute a holonomy for a
surface, and where the equation $\dt(B) = dA + A \wedge A $ comes from.
Of course
\[             dA + A \wedge A = F  \]
is just the \emph{curvature} of the connection $A$.  This is a big clue.

Suppose we are trying to compute the holonomy for a surface starting
from a $\g$-valued 1-form $A$ and an $\h$-valued 2-form $B$.  Then
following the ideas of calculus, we can try to chop the surface into
many small pieces, compute a holonomy for each one, and multiply these
together somehow.  It is easy to chop a surface into small squares.
Unfortunately, the definition of 2-category doesn't seem to know
anything about squares!  But this is not a serious problem.  For
example, we can interpret this square:
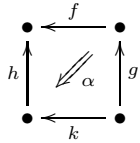
\[
  \xymatrix{
\bullet & \bullet \ar[l]_{f}="g1"  \\  
\bullet \ar[u]^{h} & \bullet \ar[l]^{k}="g2"  \ar[u]_{g}
\ar@{=>}^{\alpha} "g2"+<1.5ex,7ex>;"g2"+<-1.5ex,4ex>
}
\]
as a 2-morphism $\alpha \maps f g \To h k$.  We can then compose
a bunch of such 2-morphisms:
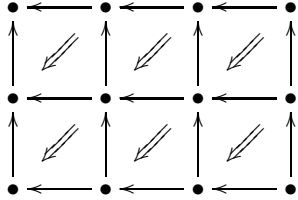
\[
\xymatrix{
\bullet & \bullet \ar[l]_{}
& \bullet \ar[l]_{}
& \bullet \ar[l]_{}
 \\  
\bullet \ar[u]^{} 
& \bullet \ar[l]^{}="g1"  \ar[u]_{}
& \bullet \ar[l]^{}="g2"  \ar[u]_{}
& \bullet \ar[l]^{}="g3"  \ar[u]_{}
\ar@{=>}^{} "g1"+<1.5ex,5.5ex>;"g1"+<-1.5ex,2.5ex>
\ar@{=>}^{} "g2"+<1.5ex,5.5ex>;"g2"+<-1.5ex,2.5ex>
\ar@{=>}^{} "g3"+<1.5ex,5.5ex>;"g3"+<-1.5ex,2.5ex>
 \\  
\bullet \ar[u]^{} 
& \bullet \ar[l]^{}="h1"  \ar[u]_{}
& \bullet \ar[l]^{}="h2"  \ar[u]_{}
& \bullet \ar[l]^{}="h3"  \ar[u]_{}
\ar@{=>}^{} "h1"+<1.5ex,5.5ex>;"h1"+<-1.5ex,2.5ex>
\ar@{=>}^{} "h2"+<1.5ex,5.5ex>;"h2"+<-1.5ex,2.5ex>
\ar@{=>}^{} "h3"+<1.5ex,5.5ex>;"h3"+<-1.5ex,2.5ex>
}
\]
with the help of a trick called `whiskering'.

Whiskering is a way to compose a 1-morphism and a 2-morphism.  Suppose we
want to compose a 2-morphism $\alpha$ and a morphism $f$ that sticks
out like a whisker on the left:
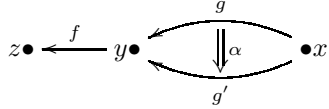
\[
\xymatrix{
  z \bullet 
& y \bullet
  \ar[l]_{f}
&&  \bullet x
  \ar@/_2.5ex/[ll]_{g}="h1"
  \ar@/^2.5ex/[ll]^{g'}="h3"
  \ar@{=>}^{\alpha} "h1"+<0ex,-2ex>;"h3"+<0ex,2ex>
}
\]
We can do this by taking the horizontal composite $1_f \circ \alpha$:
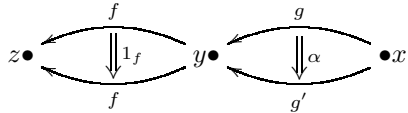
\[
\xymatrix{
 z \bullet &&
y \bullet \ar@/_2.5ex/[ll]_{f}="h1"
  \ar@/^2.5ex/[ll]^{f}="h3"
  \ar@{=>}^{1_f} "h1"+<0ex,-2ex>;"h3"+<0ex,2ex>
&& \bullet x
  \ar@/_2.5ex/[ll]_{g}="h1"
  \ar@/^2.5ex/[ll]^{g'}="h3"
  \ar@{=>}^{\alpha} "h1"+<0ex,-2ex>;"h3"+<0ex,2ex>
}
\]
We call the result $f \circ \alpha$, or $\alpha$ \define{left whiskered}
by $f$.  Similarly, if we have a whisker sticking out on the right:
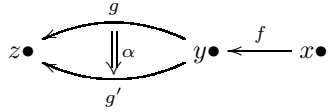
\[
\xymatrix{
z \bullet &&
y \bullet 
  \ar@/_2.5ex/[ll]_{g}="h1"
  \ar@/^2.5ex/[ll]^{g'}="h3"
  \ar@{=>}^{\alpha} "h1"+<0ex,-2ex>;"h3"+<0ex,2ex>
& x \bullet
  \ar[l]_{f}
}
\]
we can take the horizontal composite $\alpha \circ 1_f$:
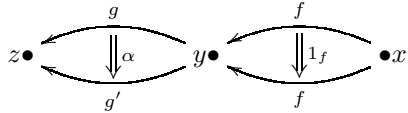
\[
\xymatrix{
z \bullet &&
y \bullet \ar@/_2.5ex/[ll]_{g}="h1"
  \ar@/^2.5ex/[ll]^{g'}="h3"
  \ar@{=>}^{\alpha} "h1"+<0ex,-2ex>;"h3"+<0ex,2ex>
&& \bullet x
  \ar@/_2.5ex/[ll]_{f}="h1"
  \ar@/^2.5ex/[ll]^{f}="h3"
  \ar@{=>}^{1_f} "h1"+<0ex,-2ex>;"h3"+<0ex,2ex>
}
\]
and call the result $\alpha \circ f$, or $\alpha$ 
\define{right whiskered} by $f$.  

With the help of whiskering, we can compose 2-morphisms shaped like
arbitrary polygons.  For example, suppose we want to horizontally
compose two squares: 
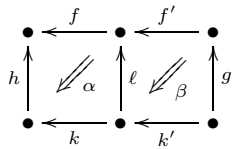
\[
\xymatrix{
\bullet 
& \bullet \ar[l]_{f}="g1"  
& \bullet \ar[l]_{f'}="h1"  
\\  
\bullet \ar[u]^{h} 
& \bullet \ar[l]^{k}="g2"  \ar[u]_{\ell}
& \bullet \ar[l]^{k'}="g3"  \ar[u]_{g}
\ar@{=>}^{\alpha} "g2"+<1.5ex,7ex>;"g2"+<-1.5ex,4ex>
\ar@{=>}^{\beta} "g3"+<1.5ex,7ex>;"g3"+<-1.5ex,4ex>
}
\]
To do this, we can left whisker $\beta$ by $f$, obtaining
this 2-morphism:
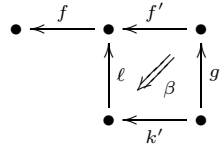
\[    f \circ \beta \maps \; f f' g \; \To \; f \ell k' \]
\[
  \xymatrix{
\bullet 
& \bullet \ar[l]_{f}="g1"  
& \bullet \ar[l]_{f'}="h1"  
\\  
& \bullet \ar[u]_{\ell}
& \bullet \ar[l]^{k'}="g3"  \ar[u]_{g}
\ar@{=>}^{\beta} "g3"+<1.5ex,7ex>;"g3"+<-1.5ex,4ex>
}
\]
Then we can right whisker $\alpha$ by $k'$, obtaining 
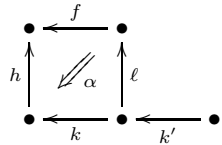
\[         \alpha \circ k' \maps \; f \ell k' \; \To \; h k k' \]
\[
\xymatrix{
\bullet 
& \bullet \ar[l]_{f}="g1"  
& 
\\  
\bullet \ar[u]^{h} 
& \bullet \ar[l]^{k}="g2"  \ar[u]_{\ell}
& \bullet \ar[l]^{k'}="g3"  
\ar@{=>}^{\alpha} "g2"+<1.5ex,7ex>;"g2"+<-1.5ex,4ex>
}
\]
Then we can vertically compose these to get the desired 2-morphism:
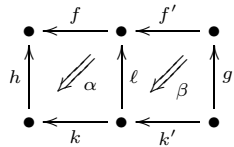
\[   (\alpha \circ k') \cdot (g \circ \beta) 
\maps \; f f' g \; \To \; h k k' \]
\[
\xymatrix{
\bullet 
& \bullet \ar[l]_{f}="g1"  
& \bullet \ar[l]_{f'}="h1"  
\\  
\bullet \ar[u]^{h} 
& \bullet \ar[l]^{k}="g2"  \ar[u]_{\ell}
& \bullet \ar[l]^{k'}="g3"  \ar[u]_{g}
\ar@{=>}^{\alpha} "g2"+<1.5ex,7ex>;"g2"+<-1.5ex,4ex>
\ar@{=>}^{\beta} "g3"+<1.5ex,7ex>;"g3"+<-1.5ex,4ex>
}
\]

The same sort of trick lets us vertically compose squares.
By iterating these procedures we can define more complicated
composites, like this:
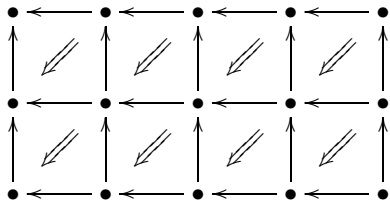
\[
\xymatrix{
\bullet & \bullet \ar[l]_{}
& \bullet \ar[l]_{}
& \bullet \ar[l]_{}
& \bullet \ar[l]_{}
 \\  
\bullet \ar[u]^{} 
& \bullet \ar[l]^{}="g1"  \ar[u]_{}
& \bullet \ar[l]^{}="g2"  \ar[u]_{}
& \bullet \ar[l]^{}="g3"  \ar[u]_{}
& \bullet \ar[l]^{}="g4"  \ar[u]_{}
\ar@{=>}^{} "g1"+<1.5ex,5.5ex>;"g1"+<-1.5ex,2.5ex>
\ar@{=>}^{} "g2"+<1.5ex,5.5ex>;"g2"+<-1.5ex,2.5ex>
\ar@{=>}^{} "g3"+<1.5ex,5.5ex>;"g3"+<-1.5ex,2.5ex>
\ar@{=>}^{} "g4"+<1.5ex,5.5ex>;"g4"+<-1.5ex,2.5ex>
 \\  
\bullet \ar[u]^{} 
& \bullet \ar[l]^{}="h1"  \ar[u]_{}
& \bullet \ar[l]^{}="h2"  \ar[u]_{}
& \bullet \ar[l]^{}="h3"  \ar[u]_{}
& \bullet \ar[l]^{}="h4"  \ar[u]_{}
\ar@{=>}^{} "h1"+<1.5ex,5.5ex>;"h1"+<-1.5ex,2.5ex>
\ar@{=>}^{} "h2"+<1.5ex,5.5ex>;"h2"+<-1.5ex,2.5ex>
\ar@{=>}^{} "h3"+<1.5ex,5.5ex>;"h3"+<-1.5ex,2.5ex>
\ar@{=>}^{} "h4"+<1.5ex,5.5ex>;"h4"+<-1.5ex,2.5ex>
}
\]
Of course, one may wonder if these more complicated composites are
unambiguously defined!  Luckily they are, thanks to associativity and
the interchange law.  This is a nontrivial result, called the `pasting
theorem' \cite{Power}.

By this method, we can reduce the task of computing $\hol(\Sigma)$ for
a large surface $\Sigma$ to the task of computing it for lots of small
squares.   Ultimately, of course, we should take a limit as the squares
become smaller and smaller.  But for our nonrigorous discussion, it is
enough to consider a very small square like this:
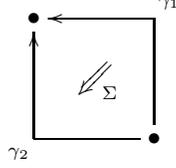
\[
\xy
(-8,8)*+{\bullet}="A";
(8,8)*{}="B";
(-8,-8)*{}="C";
(8,-8)*+{\bullet}="D";
(2,2)*{}="E";
(-2,-2)*{}="F";
"C";"D" **\dir{-};
"B";"D" **\dir{-};
{\ar@{=>}^{\scriptstyle \Sigma} "F";"E";};
{\ar@{->} "B";"A"};
{\ar@{->} "C";"A"};
(-10,-10)*{\scriptstyle \gamma_2};
(10,10)*{\scriptstyle \gamma_1};
\endxy
\]
We can think of this square as a 2-morphism 
\[   \Sigma \maps \gamma_1 \To \gamma_2 \]
where $\gamma_1$ is the path that goes up and then across,
while $\gamma_2$ goes across and then up.  We wish to compute
\[   \hol(\Sigma) \maps \hol(\gamma_1) \To \hol(\gamma_2) . \]
On the one hand, $\hol(\Sigma)$ involves the 2-form $B$.  On the
other hand, its source and target depend only on the 1-form $A$:
\[      \hol(\gamma_1) =  \poexp \left( \int_{\gamma_i} A \right) , 
\qquad 
        \hol(\gamma_2) =  \poexp \left( \int_{\gamma_2} A \right) .\]
So, $\hol(\Sigma)$ cannot have the right source and target unless $A$
and $B$ are related by an equation!

Let us try to guess this equation.  Recall from Theorem \ref{crossed_module}
that a 2-morphism $\alpha \maps g_1 \To g_2$ in $\G$ is determined by 
an element $h \in H$ with $g_2 = t(h) g_1$.
Using this, we may think of $\hol(\Sigma) \maps \hol(\gamma_1) \to 
\hol(\gamma_2)$ as determined by an element $h \in H$ with
\[        \poexp \left( \int_{\gamma_2} A \right) = 
         t(h) \; \poexp \left( \int_{\gamma_1} A \right) , \]
or in other words
\begin{equation}
\label{constraint}
       t(h) = \poexp \left( \int_{\partial \Sigma} A \right) 
\end{equation}
where the loop $\partial \Sigma = \gamma_2 \gamma_1^{-1}$ goes around
the square $\Sigma$.  For a very small square, we can approximately
compute the right hand side using Stokes' theorem:
\[       \poexp \left( \int_{\partial \Sigma} A \right) \approx
                 \exp\left( \int_\Sigma F \right) . \] 
On the other hand, there
is an obvious guess for the approximate value of $h$, which is
supposed to be built using the 2-form $B$:
\[          h \approx \exp\left( \int_\Sigma B\right) . \]
For this guess to yield Equation (\ref{constraint}), 
at least to first order in the size of our square, we need
\[    t(\exp\left( \int_\Sigma B\right)) 
\approx \exp\left( \int_\Sigma F \right) . \]
But this will be true if
\[          \dt(B) = F  .\]
And this is the equation that relates $A$ and $B$!

What have we learned here?  First, for any surface $\Sigma \maps \gamma_1
\To \gamma_2$, the holonomy $\hol(\Sigma)$ is determined by an element
$h \in H$ with 
\[        \poexp \left( \int_{\gamma_2} A \right) = 
         t(h) \; \poexp \left( \int_{\gamma_1} A \right)   \]
In the limit where $\Sigma$ is very small, this element $h$ depends 
only on $B$:
\[         h \approx \exp\left( \int_\Sigma B \right)  .\]
But for a finite-sized surface, this formula is no good, since it 
involves adding up $B$ at different points, which is not a smart
thing to do.  For a finite-sized surface, $h$ depends on $A$ as well as
$B$, since we can approximately compute $h$ by chopping this surface 
into small squares, whiskering them with paths, and composing them---and 
the holonomies along these paths are computed using $A$.  

To get the \emph{exact} holonomy over a finite-sized surface by
this method, we need to take a limit where we subdivide the surface
into ever smaller squares.   This is the Lie 2-group analogue
of a Riemann sum.  But for actual calculations, this process 
is not very convenient. More practical formulas for computing 
holonomies over surfaces can be found in the work of Schreiber and Waldorf 
\cite{SchreiberWaldorf:gerbes}, Martins and Picken \cite{MartinsPicken:2007}.

\section{Examples and Applications}
\label{examples}

Now let us give some examples of Lie 2-groups, and see what higher
gauge theory can do with these examples.  We will build these examples
using crossed modules.  Throughout what follows, $\G$ is a Lie 2-group
whose corresponding crossed module is $(G,H,t,\alpha)$.

\subsection{Shifted Abelian Groups}
\label{abelian}

Any group $G$ automatically gives a 2-group where $H$ is 
trivial.  Then higher gauge theory reduces to ordinary gauge theory.  
But to see what is \emph{new} about higher gauge theory, let us
instead suppose that $G$ is the trivial group. 
Then $t$ and $\alpha$ are forced to be trivial, and $t$ is
automatically $G$-equivariant.  On the other hand, the Peiffer identity
\[
	\alpha(t(h))h'  =  hh'h^{-1} 
\]
is not automatic: it holds if and only if $H$ is abelian!  

There is also a nice picture proof that $H$ must be abelian when
$G$ is trivial.  We simply move two elements of $H$ around each other
using the interchange law:
	
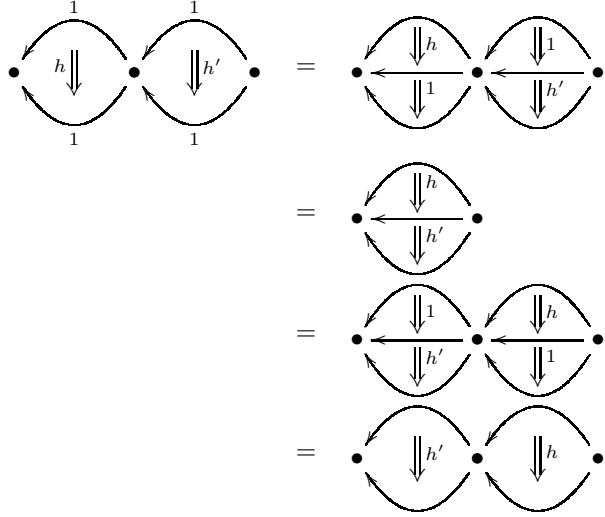
\begin{eqnarray*}
	\xy
	(-16,0)*+{\bullet}="4";
	(0,0)*+{\bullet}="6";
	{\ar@/^1.65pc/^1 "6";"4"};
	{\ar@/_1.65pc/_1 "6";"4"};
	{\ar@{=>}_<<<{h} (-8,3)*{};(-8,-3)*{}};
	(0,0)*+{\bullet}="4";
	(16,0)*+{\bullet}="6";
	{\ar@/^1.65pc/^1 "6";"4"};
	{\ar@/_1.65pc/_1 "6";"4"};
	{\ar@{=>}^<<<{h'} (8,3)*{};(8,-3)*{}};
	\endxy
	& = & 
	\xy
	(-16,0)*+{\bullet}="4";
	(0,0)*+{\bullet}="6";
	{\ar "6";"4"};
	{\ar@/^1.75pc/ "6";"4"};
	{\ar@/_1.75pc/ "6";"4"};
	{\ar@{=>}^<<<{h} (-8,6)*{};(-8,1)*{}};
	{\ar@{=>}^<<{1} (-8,-1)*{};(-8,-6)*{}};
	(0,0)*+{\bullet}="4";
	(16,0)*+{\bullet}="6";
	{\ar "6";"4"};
	{\ar@/^1.75pc/ "6";"4"};
	{\ar@/_1.75pc/ "6";"4"};
	{\ar@{=>}^<<<{1} (8,6)*{};(8,1)*{}} ;
	{\ar@{=>}^<<{h'} (8,-1)*{};(8,-6)*{}} ;
	\endxy
	\\
	& = &
	\xy 
	(-8,0)*+{\bullet}="4"; 
	(8,0)*+{\bullet}="6"; 
	{\ar@/^1.75pc/ "6";"4"}; 
	{\ar "6";"4"};
	{\ar@/_1.75pc/ "6";"4"};
	{\ar@{=>}^<<{h} (0,6)*{};(0,1)*{}} ;
	{\ar@{=>}^<<{h'} (0,-1)*{};(0,-6)*{}} ;
	\endxy
	\\
	& = &
	\xy
	(-16,0)*+{\bullet}="4";
	(0,0)*+{\bullet}="6";
	{\ar "6";"4"};
	{\ar@/^1.75pc/ "6";"4"};
	{\ar@/_1.75pc/ "6";"4"};
	{\ar@{=>}^<<<{1} (-8,6)*{};(-8,1)*{}};
	{\ar@{=>}^<<{h'} (-8,-1)*{};(-8,-6)*{}};
	(0,0)*+{\bullet}="4";
	(16,0)*+{\bullet}="6";
	{\ar "6";"4"};
	{\ar@/^1.75pc/ "6";"4"};
	{\ar@/_1.75pc/ "6";"4"};
	{\ar@{=>}^<<<{h} (8,6)*{};(8,1)*{}} ;
	{\ar@{=>}^<<{1} (8,-1)*{};(8,-6)*{}} ;
	\endxy
	\\
	& = & 
	\xy
	(-16,0)*+{\bullet}="4";
	(0,0)*+{\bullet}="6";
	{\ar@/^1.65pc/ "6";"4"};
	{\ar@/_1.65pc/ "6";"4"};
	{\ar@{=>}^<<<{h'} (-8,3)*{};(-8,-3)*{}};
	(0,0)*+{\bullet}="4";
	(16,0)*+{\bullet}="6";
	{\ar@/^1.65pc/ "6";"4"};
	{\ar@/_1.65pc/ "6";"4"};
	{\ar@{=>}^<<<{h} (8,3)*{};(8,-3)*{}};
	\endxy
	\end{eqnarray*}
As a side-benefit, we see that horizontal and vertical
composition must be equal when $G$ is trivial.  This proof is called
the `Eckmann--Hilton argument', since Eckmann and Hilton used it to
show that the second homotopy group of a space is abelian \cite{EH}.

So, we can build a 2-group where:
\begin{itemize}
\item $G$ is the trivial group,
\item $H$ is any abelian Lie group, 
\item $\alpha$ is trivial, and
\item $t$ is trivial.
\end{itemize}
This is called the \define{shifted} version of $H$, 
and denoted $\mathrm{b}H$.  

In applications to physics, we often see $H = \U(1)$.  A principal
$\mathrm{b}\U(1)$-2-bundle is usually called a \define{\boldmath
$\U(1)$ gerbe}, and a 2-connection on such a thing is usually just
called a connection.  By Theorem~\ref{thm:2-connections}, a
connection on a trivial $\U(1)$ gerbe is just an ordinary
real-valued 2-form $B$.  Its holonomy is given by:
\[
\begin{array}{rccl}
 \hol \maps& \xymatrix{ \bullet &
  \bullet \ar@/_0.5pc/[l]_{\gamma}  
} 
\quad &\mapsto & \quad 1 
\\ 
\\
\hol  \maps &
	{\xy
	(-16,0)*+{\bullet}="4";
	(0,0)*+{\bullet}="6";
	{\ar@/^1.65pc/ "6";"4"};
	{\ar@/_1.65pc/ "6";"4"};
	{\ar@{=>}^<<<{\scriptstyle \Sigma} (-8,3)*{};(-8,-3)*{}};
	\endxy}
\quad	&\mapsto & 
	\displaystyle{\exp\left( {i \int_\Sigma B} \right)} \in \U(1).
\end{array}
\]

The book by Brylinski \cite{Brylinski} gives a rather extensive
introduction to $\U(1)$ gerbes and their applications.  Murray's
theory of `bundle gerbes' gives a different viewpoint \cite{Murray,
Stevenson}.  Here let us discuss two places where $\U(1)$ gerbes
show up in physics.  One is `multisymplectic geometry'; the other is
`2-form electromagnetism'.  The two are closely related.

First, let us remember how 1-forms show up in symplectic geometry and
electromagnetism.  Suppose we have a point particle moving in some
manifold $M$.  At any time its position is a point $q \in M$ and its
momentum is a cotangent vector $p \in T^*_q M$.  As time passes, its
position and momentum trace out a curve
\[              \gamma \maps [0,1] \to T^* M  . \]
The action of this path is given by
\[ S(\gamma) = \int_\gamma \left( p_i \dot{q}^i - H(q, p) \right) \, dt \]
where $H \maps T^* M \to \R$ is the Hamiltonian.  But now suppose
the Hamiltonian is zero!  Then there is still a nontrivial action, 
due to the first term.  We can rewrite it as follows:
\[     S(\gamma) = \int_\gamma \alpha \]
where the 1-form
\[          \alpha = p_i dq^i  \]
is a canonical structure on the cotangent bundle.  We can think of 
$\alpha$ as connection on a trivial $\U(1)$-bundle over $T^* M$.
Physically, this connection describes how a quantum particle changes
phase even when the Hamiltonian is zero!  The change in phase is
computed by exponentiating the action.  So, we have:
\[     \hol(\gamma) = \exp \left( i \int_\gamma \alpha \right)  .\]

Next, suppose we carry our particle around a small loop $\gamma$ 
which bounds a disk $D$.  Then Stokes' theorem gives
\[
	S(\gamma) =  \int_\gamma \alpha
                  =  \int_D  d \alpha
\]
Here the 2-form 
\[        \omega = d \alpha = dp_i \wedge dq^i  \]
is the curvature of the connection $\alpha$.  It makes $T^*M$ into
a \define{symplectic manifold}, that is, a manifold with a
closed 2-form $\omega$ satisfying the nondegeneracy condition 
\[       \forall v \; \omega(u, v) = 0 \; \Longrightarrow \; u = 0  .\]
The subject of symplectic geometry is vast and deep, but sometimes
this simple point is neglected: the symplectic structure describes
the \emph{change in phase} of a quantum particle as we move it around a
loop:
\[         \hol(\gamma) =  \exp \left( i \int_D \omega \right) .\] 
Perhaps this justifies calling a symplectic manifold a `phase
space', though historically this seems to be just a coincidence.

It may seem strange to talk about a quantum particle tracing out a
loop in phase space, since in quantum mechanics we cannot
simultaneously know a particle's position and momentum.  However,
there is a long line of work, beginning with Feynman, which computes
time evolution by an integral over paths in phase space \cite{DMN}.
This idea is also implicit in geometric quantization, where the first
step is to equip the phase space with a principal $\U(1)$-bundle
having a connection whose curvature is the symplectic structure.  (Our
discussion so far is limited to \emph{trivial} bundles, but everything
we say generalizes to the nontrivial case.)

Next, consider a charged particle in an electromagnetic field.
Suppose that we can describe the electromagnetic field using a vector
potential $A$ which is a connection on trivial $\U(1)$ bundles over
$M$.  Then we can pull $A$ back via the projection $\pi \maps T^\ast M
\to M$, obtaining a 2-form $\pi^*\!A$ on phase space.  In the absence
of any other Hamiltonian, the particle's action as we move it
along a path $\gamma$ in phase space will be
\[    S(\gamma) =   \int_\gamma \alpha + e\,\pi^*\!A  \]
if the particle has charge $e$.  In short, \textit{the
electromagnetic field changes the connection on phase space}
from $\alpha$ to $\alpha + e\, \pi^*\!A$.  Similarly, when the path $\gamma$ 
is a loop bounding a disk $D$, we have
\[    S(\gamma) = \int_D \omega + e\,\pi^*\!F   \]
where $F = dA$ is the electromagnetic field strength.  So,
electromagnetism also changes the symplectic structure on phase
space from $\omega$ to $\omega + e \, \pi^*\!F$.  For more on this, see
Guillemin and Sternberg \cite{GS}, who also treat the case of
nonabelian gauge fields.

All of this has an analog where particles are replaced by strings.  It
has been known for some time that just as the electromagnetic vector
potential naturally couples to point particles, there is a 2-form $B$
called the Kalb--Ramond field which naturally couples to strings.  The
action for this coupling is obtained simply by integrating $B$ over
the string worldsheet.  In 1986, Gawedski \cite{Gawedski} showed that
the $B$ field should be seen as a connection on a $\U(1)$ gerbe.
Later Freed and Witten \cite{FreedWitten} showed this viewpoint was
crucial for understanding anomaly cancellation.  However, these
authors did not actually use the word `gerbe'.  The role of gerbes was
later made explicit by Carey, Johnson and Murray \cite{CJM}, and even
more so by Gawedski and Reis \cite{GawedskiReis}.

In short, electromagnetism has a `higher version'.  What about
symplectic geometry?  This also has a higher version, which dates back to
1935 work by DeDonder \cite{DeDonder} and Weyl \cite{Weyl}.  The idea
here is that an $n$-dimensional classical field theory has a kind of
finite-dimensional phase space equipped with a closed $(n+2)$-form
$\omega$ which is \define{nondegenerate} in the following sense:
\[       \forall v_1,\dots,v_{n+1} \; \; \omega(u, v_1, \dots, v_n) = 0 
\quad \Longrightarrow \quad u = 0  .\]
Such a $n$-form is called a \define{multisymplectic structure}, or
more specifically, an \define{$n$-plectic structure} For a nice
introduction to multisymplectic geometry, see the paper by Gotay,
Isenberg, Marsden, and Montgomery \cite{GIMM}.

The link between multisymplectic geometry and higher electromagnetism
was made in a paper by Baez, Hoffnung and Rogers
\cite{BaezHoffnungRogers}.  Everything is closely analogous to the
story for point particles.  For a classical bosonic string propagating
on Minkowski spacetime of any dimension, say $M$, there is a
finite-dimensional manifold $X$ which serves as a kind of `phase
space' for the string.  There is a projection $\pi \maps X \to M$, and
there is a god-given way to take any map from the string's worldsheet
to $M$ and lift it to an embedding of the worldsheet in $X$.  So, let
us write $\Sigma$ for the string worldsheet considered as a surface in
$X$.

The phase space $X$ is equipped with a 2-plectic structure: that is, a
closed nondegenerate 3-form, say $\omega$.  But in fact, $\omega = d
\alpha$ for some 2-form $\alpha$.  Even when the string's Hamiltonian
is zero, there is a term in the action of the string coming from the
integral of $\alpha$:
\[          S(\Sigma) = \int_{\Sigma} \alpha .\]
We may also consider a charged string coupled to a Kalb--Ramond field.
This begins life as a 2-form $B$ on $M$, but we may pull it back to a 
2-form $\pi^*\!B$ on $X$, and then
\[    S(\Sigma) =   \int_\Sigma \alpha + e\,\pi^*\!B.  \]
In particular, suppose $\Sigma$ is a 2-sphere bounding a 3-ball 
$D$ in $X$.  Then by Stokes' theorem we have
\[    S(\Sigma) =   \int_D \omega + e\,\pi^*\!Z  \]
where the 3-form
\[         Z = dB  \]
is the Kalb--Ramond analog of the electromagnetic field strength,
and $e$ is the string's charge.  (The Kalb--Ramond field strength 
is usually called `$H$' in the physics literature, but that conflicts 
with our usage of $H$ to mean a Lie group, so we shall call it `$Z$'.) 

In summary: \emph{the Kalb--Ramond field modifies the 2-plectic 
structure on the phase space of the string}.
The reader will note that we have coyly refused to describe the phase
space $X$ or its 2-form $\alpha$.  For this, see the paper by Baez,
Hoffnung and Rogers \cite{BaezHoffnungRogers}.  In this paper, we
explain how the usual dynamics of a classical bosonic string coupled
to a Kalb--Ramond field can be described using multisymplectic
geometry.  We also explain how to generalize Poisson brackets from
symplectic geometry to multisymplectic geometry.  Just as Poisson
brackets in symplectic geometry make the functions on phase space into
a Lie algebra, Poisson brackets in multisymplectic geometry give rise
to a `Lie 2-algebra'.  Lie 2-algebras are also important in higher
gauge theory in the same way that Lie algebras are important for gauge
theory.  Indeed, the `string 2-group' described in Section
\ref{string} was constructed only after its Lie 2-algebra was found
\cite{BaezCrans}.  Later, this Lie 2-algebra was seen to arise
naturally from multisymplectic geometry \cite{BaezRogers}.

\subsection{The Poincar\'e 2-Group}
\label{poincare}

Suppose we have a representation $\alpha$ of a Lie group $G$ on a
finite-dimensional vector space $H$.  We can regard $H$ as an abelian
Lie group with addition as the group operation.  This lets us
regard $\alpha$ as an action of $G$ on this abelian Lie group.
So, we can build a 2-group $\G$ where:
\begin{itemize}
\item $G$ is any Lie group,
\item $H$ is any vector space,
\item $\alpha$ is the representation of $G$ on $H$, and
\item $t$ is trivial.
\end{itemize}
In particular, note that the Peiffer identity holds.  In this way, we
see that any group representation gives a crossed module---so group
representations are secretly 2-groups!

For example, if we let $G$ be the Lorentz group and let $\alpha$ be
its obvious representation on $\R^4$:
\[ G = \SO(3,1) \]
\[ H = \R^4 \]
we obtain the so-called \define{Poincar\'e 2-group}, which has the
Lorentz group as its group of morphisms, and the Poincar\'e group as
its group of 2-morphisms \cite{BaezLauda:2groups}.

What is the Poincar\'e 2-group good for?  It is not clear, but there
are some clues.  Just as we can study representations of groups on
vector spaces, we can study representations of 2-groups on `2-vector
spaces' \cite{BBFW, BarrettMackaay,CraneYetter,Elgueta2}.  The
representations of a group are the objects of a category, and this
sort of category can be used to build `spin foam models' of
background-free quantum field theories \cite{Baez:spinfoam}.  This
endeavor has been most successful with 3d quantum gravity
\cite{FreidelLouapre}, but everyone working on this subject dreams
of doing something similar for 4d quantum gravity \cite{Rovelli}.
Going from groups to 2-groups boosts the dimension of everything:
the representations of a 2-group are the objects of a
\emph{2-category}, and Crane and Sheppeard outlined a program for
building a 4-dimensional spin foam model starting from the 2-category of
representations of the Poincar\'e 2-group \cite{CraneSheppeard}.

Crane and Sheppeard hoped their model would be related to quantum
gravity in 4 spacetime dimensions.  This has not come to pass, at least
not yet---but this spin foam model \textit{does} have interesting
connections to 4d physics.  The spin foam model of 3d quantum gravity
automatically includes point particles, and Baratin and Freidel have
shown that it reduces to the usual theory of Feynman diagrams in 3d
Minkowski spacetime in the limit where the gravitational constant
$G_{\textrm{{\tiny Newton}}}$ goes to zero \cite{BaratinFreidel:3d}.
This line of thought led Baratin and Freidel to construct a spin foam
model that is equivalent to the usual theory of Feynman diagrams in 4d
Minkowski spacetime \cite{BaratinFreidel:4d}.  At first the
mathematics underlying this model was a bit mysterious---but it now
seems clear that this model is based on the representation theory of
the Poincar\'e 2-group!  For a preliminary report on this fascinating
research, see the paper by Baratin and Wise \cite{BaratinWise}.

In short, it appears that the 2-category of representations of the
Poincar\'e 2-group gives a spin foam description of quantum field
theory on 4d Minkowski spacetime.  Unfortunately, while spin foam
models in 3 dimensions can be obtained by quantizing gauge theories,
we do not see how to obtain this 4d spin foam model by quantizing a
higher gauge theory.  Indeed, we know of no classical field theory in
4 dimensions whose solutions are 2-connections on a principal
$\G$-2-bundle where $\G$ is the Poincar\'e 2-group.

However, if we replace the Poincar\'e 2-group by a closely related
2-group, this puzzle \emph{does} have a nice solution.  Namely, if
we take 
\[ G = \SO(3,1) \]
\[ H = \so(3,1) \]
and take $\alpha$ to be the adjoint representation, we obtain the
`tangent 2-group' of the Lorentz group.  As we shall see,
2-connections for this 2-group arise naturally as solutions of a 
4d field theory called `topological gravity'.  

\subsection{Tangent 2-Groups}
\label{tangent}

We have seen that any group representation gives a 2-group.  But any
Lie group $G$ has a representation on its own Lie algebra: the adjoint
representation.  This lets us build a 2-group from the crossed module
where:
\begin{itemize}
\item $G$ is any Lie group,
\item $H$ is $\g$ regarded as a vector space and thus an abelian
Lie group, 
\item $\alpha$ is the adjoint representation, and
\item $t$ is trivial.
\end{itemize}
We call this the \define{tangent 2-group} $\T G$ of the Lie group $G$.
Why?  We have already seen that for any Lie 2-group, the group of all
2-morphisms is the semidirect product $G \ltimes H$.  In the case at
hand, this semidirect product is just $G \ltimes \g$, with $G$ acting
on $\g$ via the adjoint representation.  But as a manifold, this
semidirect product is nothing other than the tangent bundle $TG$ of
the Lie group $G$.  So, the tangent bundle $TG$ becomes a group, and
this is the group of 2-morphisms of $\T G$.

By Theorem~\ref{thm:2-connections}, a 2-connection on a trivial $\T
G$-2-bundle consists of a $\g$-valued 1-form $A$ and a $\g$-valued 
2-form $B$ such that the curvature $F = dA + A \wedge A$ satisfies
\[   F = 0, \]
since $\dt(B) = 0$ in this case.  Where can we find such 2-connections?
We can find them as solutions of a field theory called 4-dimensional
$BF$ theory!

$BF$ theory is a classical field theory that works in any dimension.
So, take an $n$-dimensional oriented manifold $M$ as our spacetime.
The fields in $BF$ theory are a connection $A$ on the trivial
principal $G$-bundle over $M$, together with a $\g$-valued
$(n-2)$-form $B$.  The action is given by
\[  S(A,B) = \int_M \tr(B \wedge F) .\]
Setting the variation of this action equal to zero, we obtain the
following field equations:
\[ dB + [A,B] = 0, \quad F = 0. \]
In dimension 4, $B$ is a $\g$-valued 2-form---and thanks to the
second equation, $A$ and $B$ fit together to define a 
2-connection on the trivial $\T G$-2-bundle over $M$.  

It may seem dull to study a gauge theory where the equations of motion
imply the connection is flat.  But there is still room for some fun.
We see this already in 3-dimensional $BF$ theory, where $B$ is a
$\g$-valued 1-form rather than a 2-form.  This lets us package $A$ 
and $B$ into a connection on the trivial $TG$-bundle over $M$.  The 
field equations
\[ dB + [A,B] = 0, \quad F = 0 \]
then say precisely that this connection is flat.

When the group $G$ is the Lorentz group $\SO(2,1)$, $TG$ is the
corresponding Poincar\'e group.  With this choice of $G$, 3d $BF$
theory is a version of 3d general relativity.  In 3 dimensions, unlike
the more physical 4d case, the equations of general relativity say
that spacetime is \emph{flat} in the absence of matter.  And at first
glance, 3d $BF$ theory only describes general relativity without
matter.  After all, its solutions are flat connections.

Nonetheless, we can consider 3d $BF$ theory on a manifold from which
the worldline of a point particle has been removed.  In the Bohm--Aharonov
effect, if we carry a charged object around a solenoid, we obtain a
nontrivial phase even though the $\U(1)$ connection $A$ is flat outside
the solenoid.  Similarly, in 3d $BF$ theory, the connection $(A,B)$ will 
be flat away from the particle's worldline, but it can have a nontrivial 
holonomy around a loop $\gamma$ that encircles the worldline:
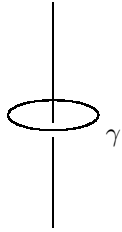
\[
\xy
(0,0)*\ellipse(6,2){-};
(8,-3)*{\gamma};
(0,-15)*{};(0,-3)*{} **\dir{-};
(0,-1)*{};(0,15)*{} **\dir{-};
\endxy
\]
This holonomy says what happens when we parallel transport an object
around our point particle.   The holonomy is an element of the Poincar\'e
group.  Its conjugacy class describes the \emph{mass} and \emph{spin} 
of our particle.  So, massive spinning point particles are lurking 
in the formalism of 3d $BF$ theory!

Even better, this theory predicts an upper bound on the particle's
mass, roughly the Planck mass.  This is true even classically.  This
may seem strange, but unlike in 4 dimensions, where we need $c$,
$G_{\textrm{{\tiny Newton}}}$ and $\hbar$ to build a quantity with
dimensions of length, in 3-dimensional spacetime we can do this using
only $c$ and $G_{\textrm{{\tiny Newton}}}$.  So, ironically, the
`Planck mass' does not depend on Planck's constant.

Furthermore, in this theory, particles have `exotic statistics',
meaning that the interchange of identical particles is 
governed by the braid group instead of the symmetric group.  
Particles with exotic statistics are also known as `anyons'.  
In the simplest examples, the anyons in 3d gravity reduce to 
bosons or fermions in the $G_{\textrm{{\tiny Newton}}} \to 0$ 
limit.

There is thus a wealth of interesting phenomena to be studied
in 3d $BF$ theory.  See the paper by Baez, Crans and Wise
\cite{BaezCransWise} for a quick overview, and the work of 
Freidel, Louapre and Baratin for a deep treatment of the details
\cite{BaratinFreidel:3d,FreidelLouapre}.

The case of 4d $BF$ theory is just as interesting, and not as fully
explored.  In this case the field equations imply that $A$
and $B$ define a 2-connection on the trivial $\T G$-2-bundle over $M$.
But in fact they say more: they say precisely that this 2-connection
is \define{flat}.  By this we mean two things.  First, the holonomy
$\hol(\gamma)$ along a path $\gamma$ does not change when we change
this path by a homotopy.  Second, the holonomy $\hol(\Sigma)$ along a
surface $\Sigma$ does not change when we change this surface by a
homotopy.  The first fact here is equivalent to the equation $F = 0$.
The second is equivalent to the equation $dB + [A,B] = 0$.

When the group $G$ is the Lorentz group $\SO(3,1)$, 4d $BF$ theory is
sometimes called `topological gravity'.  We can think of it as a
simplified version of general relativity that acts more like gravity
in 3 dimensions.  In particular, we can copy what we did in 3
dimensions, and consider 4d $BF$ theory on a manifold from which the
worldlines of particles \emph{and the worldsheets of strings} have
been removed.  Some of what we will do here works for more general
groups $G$, but let us take $G = \SO(3,1)$ just to be specific.

First consider strings.  Take a 2-dimensional manifold $X$ embedded
in a 4-dimensional manifold $M$, and think of $X$ as the
worldsheet of a string.  Suppose we can find a small loop $\gamma$ that 
encircles $X$ in such a way that $\gamma$ is contractible in $M$ 
but not in $M - X$.  If we do 4d $BF$ theory on the spacetime $M - X$, the
holonomy
\[    \hol(\gamma) \in \SO(3,1)  \] 
will not change when we apply a homotopy to $\gamma$.  This holonomy
describes the `mass density' of our string \cite{BaezCransWise,
Balachandran2}.

Next, consider particles.  Take a curve $C$ embedded in $M$, and think
of $C$ as the worldline of a particle.  Suppose we can find a small 2-sphere
$\Sigma$ in $M - C$ that is contractible in $M$ but not $M - C$.  We
can think of this 2-sphere as a 2-morphism $\Sigma \maps 1_x \To 1_x$
in the path 2-groupoid of $M$.  If we do 4d $BF$ theory on the
spacetime $M - C$, the holonomy
\[     \hol(\Sigma) \in \so(3,1) \] 
will not change when we apply a homotopy to $\Sigma$.  So, this
holonomy describes some information about the particle---but so
far as we know, the physical meaning of this information has not been
worked out.  

What if we had a field theory whose solutions were flat 2-connections
for the Poincar\'e 2-group?  Then we would have 
\[     \hol(\Sigma) \in \R^4 \] 
and there would be a tempting interpretation of this quantity: namely,
as the energy-momentum of our point particle.  So, the puzzle posed at
the end of the previous section is a tantalizing one.

One may rightly ask if the `strings' described above bear any relation
to those of string theory.  If they are merely surfaces cut out of
spacetime, they lack the dynamical degrees of freedom normally
associated to a string. Certainly they do not have an action
proportional to their surface area, as for the Polyakov string.
Indeed, one may ask if `area' even makes sense in 4d $BF$ theory.
After all, there is no metric on spacetime: the closest substitute is
the $\so(3,1)$-valued 2-form $B$.

Some of these problems may have solutions.  For starters, when
we remove a surface $X$ from our 4-manifold $M$, the action 
\[ S(A,B) = \int_{M-X} \tr(B \wedge F) \]
is no longer gauge-invariant: a gauge transformation changes the
action by a boundary term which is an integral over $X$.  We can
remedy this by introducing fields that live on $X$, and adding a term
to the action which is an integral over $X$ involving these fields.
There are a number of ways to do this \cite{BaezPerez, Fairbairn,
GirelliPfeifferPopescu, FNS, MP}.  For some, the integral over $X$ is
proportional to the area of the string worldsheet in the special case
where the $B$ field arises from a cotetrad (that is, an $\R^4$-valued
1-form) as follows:
\[           B = e \wedge e  \]
where we use the isomorphism $\Lambda^2 \R^4 \iso \so(3,1)$.  
In this case there is close relation to the Nambu--Goto string,
which has been carefully examined by Fairbairn, Noui and Sardelli
\cite{FNS}.  

This is especially intriguing because when $B$ takes
the above form, the $BF$ action becomes the usual Palatini action
for general relativity:
\[  S(A,e) = \int_M \tr(e \wedge e \wedge F) \]
where `$\tr$' is a suitable nondegenerate bilinear form on $\so(3,1)$.
Unfortunately, solutions of Palatini gravity typically \emph{fail} to
obey the condition $\dt(B) = F$ when we take $B = e \wedge e$.  So, we
cannot construct 2-connections in the sense of Theorem
\ref{thm:2-connections} from these solutions!  If we want to treat
general relativity in 4 dimensions as a higher gauge theory, we need
other ideas.  We describe two possibilities at the end of Section
\ref{automorphism}.

\subsection{Inner Automorphism 2-Groups}
\label{inner}

There is also a Lie 2-group where:
\begin{itemize}
\item $G$ is any Lie group,
\item $H = G$,
\item $t$ is the identity map,
\item $\alpha$ is conjugation:
\[ \alpha(g) h = ghg^{-1}. \]
\end{itemize}
Following Roberts and Schreiber \cite{RobertsSchreiber} we call this
the \define{inner automorphism 2-group} of $G$, and denote it by
$\INN(G)$.  We explain this terminology in the next
section.  

A 2-connection on the trivial $\INN(G)$-2-bundle over a manifold
consists of a $\g$-valued 1-form $A$ and a $\g$-valued 2-form $B$ such
that
\[  B = F  \]
since $\dt$ is now the identity.  Intriguingly, 2-connections of this
sort show up as solutions of a slight variant of 4d $BF$ theory.  In a
move that he later called his biggest blunder, Einstein took general
relativity and threw an extra term into the equations: a `cosmological
constant' term, which gives the vacuum nonzero energy.  We can do the
same for topological gravity, or indeed 4d $BF$ theory for any group
$G$.  After all, what counts as a blunder for Einstein might count
as a good idea for lesser mortals such as ourselves.

So, fix a $4$-dimensional oriented manifold $M$ as our spacetime.  As
in ordinary $BF$ theory, take the fields to be a connection $A$ on the
trivial principal $G$-bundle over $M$, together with a $\g$-valued
$2$-form $B$.  The action for $BF$ theory `with cosmological constant'
is defined to be
\[  S(A,B) = \int_M \tr(B \wedge F - \frac{\lambda}{2} B \wedge B). \]
Setting the variation of the action equal to zero, we obtain these
field equations:
\[ dB + [A,B] = 0 , \quad F = \lambda B . \]
When $\lambda = 0$, these are just the equations we saw in the
previous section.  But let us consider the case $\lambda \ne 0$.  Then
these equations have a drastically different character!  The Bianchi
identity $dF + [A,F] = 0$, together with $F = \lambda B$,
automatically implies that $dB + [A,B] = 0$.  So, to get a solution of
this theory we simply take any connection $A$, compute its curvature
$F$ and set $B = F/\lambda$.

This may seem boring: a field theory where any connection is a
solution.  But in fact it has an interesting relation to higher gauge
theory.  To see this, it helps to change variables and work with the
field $\beta = \lambda B$.  Then the field equations become
\[ d \beta + [A,\beta] = 0 , \quad F = \beta. \]
Any solution of these equations gives a 2-connection on the
trivial principal $\INN(G)$-2-bundle over $M$!

There is also a tantalizing relation to the cosmological
constant in general relativity.  If the $B$ field arises
from a cotetrad as explained in the previous section:
\[           B = e \wedge e , \]
then the above action becomes 
\[  S = \int_M \tr(e \wedge e \wedge F - \frac{\lambda}{2} e \wedge
e \wedge e \wedge e). \]
When we choose the bilinear form `$\tr$' correctly, 
this is the action for general relativity with a cosmological 
constant proportional to $\lambda$.  

There is some evidence \cite{Baez:BF} that $BF$ theory with nonzero
cosmological constant can be quantized to obtain the so-called
Crane--Yetter model \cite{CKY, CY}, which is a spin foam model based
on the category of representations of the quantum group associated to
$G$.  Indeed, in some circles this is taken almost as an article of
faith.  But a rigorous argument, or even a fully convincing argument,
seems to be missing.  So, this issue deserves more study.

The $\lambda \to 0$ limit of $BF$ theory is fascinating but highly
singular, since for $\lambda \ne 0$ a solution is just a connection
$A$, while for $\lambda = 0$ a solution is a flat connection $A$
together with a $B$ field such that $dB + [A,B] = 0$.  At least in some
rough intuitive sense, as $\lambda \to 0$ the group $H$ in the crossed
module corresponding to $\INN(G)$ `expands and flattens out'
from the group $G$ to its tangent space $\g$.  Thus, $\INN(G)$
degenerates to the tangent 2-group $\T G$.  It would be nice to make
this precise using a 2-group version of the theory of group contractions.

\subsection{Automorphism 2-Groups}
\label{automorphism}

The inner automorphism group of the previous section is
closely related to the \define{automorphism 2-group} $\AUT(H)$, 
defined using the crossed module where:
\begin{itemize}
\item $G = \Aut(H)$,
\item $H$ is any Lie group,
\item $t \maps H \to \Aut(H)$ sends any group element to the operation
of conjugating by that element,
\item $\alpha \maps \Aut(H) \to \Aut(H)$ is the identity.
\end{itemize}
We use the term `automorphism 2-group' because $\AUT(H)$ really is the
2-group of symmetries of $H$.  Lie groups form a 2-category, any object
in a 2-category has a 2-group of symmetries, and the 2-group of 
symmetries of $H$ is naturally a Lie 2-group, which is none other than
$\AUT(H)$.  See \cite{BaezLauda:2groups} for details.

A principal $\AUT(H)$-2-bundle is usually called a \define{nonabelian 
gerbe} \cite{Breen:2006}.  Nonabelian gerbes are a major test case 
for ideas in higher gauge theory.  Indeed, almost the whole 
formalism of 2-connections was worked out first for nonabelian 
gerbes by Breen and Messing \cite{BreenMessing}.  The one aspect 
they did not consider is the one we have focused on here: parallel 
transport.  Thus, they did not impose the equation $\dt(B) = F$, 
which we need to obtain holonomies satisfying the conditions of 
Theorem \ref{thm:2-connections}.  Nonetheless, the quantity 
$F - \dt(B)$ plays an important role in Breen and Messing's 
formalism: they call it the \define{fake curvature}.  Generalizing
their ideas slightly, for any Lie 2-group $\G$, we may define a 
\define{connection} on a trivial principal $\G$-2-bundle to be a 
pair consisting of a $\g$-valued 1-form $A$ and an $\h$-valued 2-form.  
A 2-connection is then a connection with vanishing fake curvature.

The relation between the automorphism 2-group and the inner
automorphism 2-group is nicely explained in the work of Roberts and
Schreiber \cite{RobertsSchreiber}.   As they discuss, for any group $G$
there is an exact sequence of 2-groups
\[  
1 \to \ZZ(G) \to \INN(G) \to \AUT(G) \to \OUT(G) \to 1
\]
where $\ZZ(G)$ is the center of $G$ and $\OUT(G)$ is the group of
outer automorphisms of $G$, both regarded as 2-groups with only
identity 2-morphisms.  

Roberts and Schreiber go on to consider an analogous sequence of
\emph{3-groups} constructed starting from a 2-group.  Among these, the
`inner automorphism 3-group' $\INN(\G)$ of a 2-group $\G$ plays a
special role \cite{Schreiber:cafe}.  The reason is that any connection
on a principal $\G$-2-bundle, not necessarily obeying $\dt(B) = F$,
gives a flat 3-connection on a principal $\INN(\G)$-3-bundle!  This in
turn allows us to define a version of parallel transport for
particles, strings \emph{and 2-branes}.

This may give a way to understand general relativity in terms of
higher gauge theory.  As we have already seen in Section
\ref{tangent}, Palatini gravity in 4d spacetime involves an
$\so(3,1)$-valued 1-form $A$ and an $\so(3,1)$-valued 2-form $B = e
\wedge e$.  This is precisely the data we expect for a connection on a
principal $\G$-2-bundle where $\G$ is the tangent 2-group of the
Lorentz group.  Typically this connection fails to obey the equation
$\dt(B) = F$.  So, it is not a 2-connection.  But, it gives a flat
3-connection on an $\INN(\T\SO(3,1))$-3-bundle.  So, we may optimistically
call $\INN(\T\SO(3,1))$ the \define{gravity 3-group}.

Does the gravity 3-group actually shed any light on general relativity?
The work of Martins and Picken \cite{MartinsPicken:2009} establishes a
useful framework for studying these issues.  They define a path
3-groupoid $\P_3(M)$ for a smooth manifold $M$.  Given a Lie 3-group
${\bf G}$, they describe 3-connections on the trivial ${\bf
G}$-3-bundle over $M$ as 3-functors
\[                   \hol \maps \P_3(M) \to {\bf G}  \]
Moreover, they show how to construct these functors from a 1-form, a
2-form, and a 3-form taking values in three Lie algebras associated to
${\bf G}$.  In the case where ${\bf G} = \INN(\T\SO(3,1))$ and 
$\hol$ is a flat 3-connection, this data reduces to an
$\so(3,1)$-valued 1-form $A$ and an $\so(3,1)$-valued 2-form $B$.

\subsection{String 2-Groups}
\label{string}

The Lie 2-groups discussed so far are easy to construct.  The string
2-group is considerably more subtle.  Ultimately it forces upon us
a deeper conception of what a Lie 2-group really is, and a more
sophisticated approach to higher gauge theory.  Treated in proper
detail, these topics would carry us far beyond the limits of this
quick introduction.  But it would be a shame not to mention them at
all.

Suppose we have a central extension of a Lie group $G$ by an
abelian Lie group $A$.  In other words,
suppose we have a short exact sequence of Lie groups
\[   1 \to A \to H \stackrel{t}{\to} G \to 1 \]
where the image of $A$ lies in the center of $H$.  
Then we can construct an action $\alpha$ of $G$ on $H$ as follows.  
The map $t \maps H \to G$ describes $H$ as a fiber bundle over $G$,
so choose a \define{section} of this bundle: that is, a function 
$s \maps G \to H$ with $t(s(g)) = g$, not necessarily a homomorphism.  
Then set
\[      \alpha(g) h = s(g) h s(g)^{-1}  .\]
Since $A$ is included in the center of $H$, $\alpha$ is independent of
the choice of $s$.  Thanks to this, we do not need a global smooth
section $s$ to check that $\alpha(g)$ depends smoothly on $g$: it
suffices that there exist a local smooth section in a
neighborhood of each $g \in G$, and indeed this is always true. We can
use these local sections to define $\alpha$ globally, since they must
give the same $\alpha$ on overlapping neighborhoods.

Given all this, we can check that $t$ is $G$-equivariant and that
the Peiffer identity holds.  So, we obtain a Lie 2-group where:
\begin{itemize}
\item $G$ is any Lie group,
\item $H$ is any Lie group,
\item $t \maps H \to G$ makes $H$ into a central extension of $G$,
\item $\alpha$ is given by $\alpha(g) h = s(g) h s(g)^{-1}$ where
$s \maps G \to H$ is any section.
\end{itemize}
We call this the \define{central extension 2-group} 
$\CC(H \stackrel{t}{\to} G)$.  

To get concrete examples, we need examples of central extensions. 
For any choice of $G$ and $A$, we can always take $H = G \times A$ and
use the `trivial' central extension
\[   1 \to A \to A \times G \to G \to 1 .\]
For more interesting examples, we need nontrivial central
extensions.  These tend to arise from problems in quantization.  For
example, suppose $V$ is a finite-dimensional 
\define{symplectic vector space}: that is, a vector space
equipped with a nondegenerate antisymmetric bilinear form
\[     \omega \maps V \times V \to \R .\]
Then we can make $H = V \times \R$ into a Lie group
called the \define{Heisenberg group}, with the product
\[    (u,a)(v,b) = (u+v, a + b + \omega(u,v)) .\]
The Heisenberg group plays a fundamental role in quantum mechanics,
because we can think of $V$ as the phase space of a classical point
particle.  If we let $G$ stand for $V$ regarded as an abelian Lie group, 
then elements of $G$ describe translations in phase space: that is,
translations of both position and momentum.  The Heisenberg group $H$
describes how these translations commute only `up to a phase'
when we take quantum mechanics into account: the phase is given by
$\exp(i \omega(u,v))$.  There is a homomorphism $t \maps H \to G$
that forgets this phase information, given by
\[    t(u,a) = u .\]
This exhibits $H$ as a central extension of $G$.  We thus
obtain a central extension 2-group $\CC(H \stackto{t} G)$, called the
{\bf Heisenberg 2-group} of the symplectic vector space $V$.

The applications of Heisenberg 2-groups seem largely unexplored, and
should be worth studying.  So far, much more work has been put into
understanding 2-groups arising from central extensions of loop groups.
The reason is that central extensions of loop groups play a basic role
in string theory and conformal field theory, as nicely explained by
Pressley and Segal \cite{PressleySegal:1986}.

Suppose that $G$ is a connected and simply-connected compact simple Lie 
group $G$.   Define the \define{loop group} $\Omega G$ to be
the set of all smooth paths $\gamma \maps [0,1] \to G$ that start and
end at the identity of $G$.  This becomes a group under pointwise
multiplication, and in fact it is a kind of infinite-dimensional Lie
group \cite{Milnor}. 

For each integer $k$, called the \define{level}, the loop group has a
central extension
\[           1 \to \U(1) \to
\widehat{\Omega_k G} \stackto{t} \Omega G \to 1 .  \]
These extensions are all different, and all nontrivial except for $k =
0$.  In physics, they arise because the 2d gauge theory called the
Wess--Zumino--Witten model has an `anomaly'.  The loop group $\Omega
G$ acts as gauge transformations in the classical version of this
theory.  However, when we quantize the theory, we obtain a
representation of $\Omega G$ only `up to a phase'---that is, a
projective representation.  This can be understood as an honest
representation of the central extension $\widehat{\Omega_k G}$, where
the integer $k$ appears in the Lagrangian for the Wess--Zumino--Witten
model.

Starting from this central extension we can construct a central
extension 2-group called the \define{\boldmath level-$k$ loop
2-group} of $G$, ${\cal L}_k(G)$.  This is an infinite-dimensional Lie
2-group, meaning that it comes from a crossed module where the groups
involved are infinite-dimensional Lie groups, and all the maps are
smooth.  Moreover, it fits into an exact sequence
\[           1 \to \mathcal{L}_k(G) \to \STRING_k(G) 
\stackto{} G \to 1   \]
where the middle term, the \textbf{level-\textit{k} string 2-group} of
$G$, has very interesting properties \cite{BCSS}.  

Since the string 2-group $\STRING_k(G)$ is an infinite-dimensional
Lie 2-group, it is a topological 2-group.  There is a way to take any
topological 2-group and squash it down to a topological group
\cite{BCSS,BaezStevenson}.  Applying this trick to $\STRING_k(G)$ when
$k = 1$, we obtain a topological group whose homotopy groups match
those of $G$---except for the third homotopy group, which has been
made trivial.  In the special case where $G = \mathrm{Spin}(n)$, this
topological group is called the `string group', since to consistently
define superstrings propagating on a spin manifold, we must reduce its
structure group from $\mathrm{Spin}(n)$ to this group
\cite{Witten:1988}.  The string group also plays a role in Stolz and
Teichner's work on elliptic cohomology, which involves a notion of
parallel transport over surfaces \cite{StolzTeichner}.  There is a lot
of sophisticated mathematics involved here, but ultimately much of it
should arise from the way string 2-groups are involved in the parallel
transport of strings!  The work of Sati, Schreiber and Stasheff
\cite{SSS} provides good evidence for this, as does the work of
Waldorf \cite{Waldorf}.

In fact, the string Lie 2-group had lived through many previous 
incarnations before being constructed as an infinite-dimensional 
Lie 2-group.  Brylinski and McLaughlin \cite{BrylinskiMcLaughlin}
thought of it as a $\U(1)$ gerbe over the group $G$.  The fact 
that this gerbe is `multiplicative' makes it something like a 
group in its own right \cite{Brylinski:2000}.  This viewpoint has
also been explored by Murray and Stevenson \cite{MurrayStevenson}.

Later, Baez and Crans \cite{BaezCrans} constructed a Lie 2-algebra 
$\sstring_k(\g)$ corresponding to the string Lie 2-group.  For 
pedagogical purposes, our discussion of Lie 2-groups has focused 
solely on `strict' 2-groups, where the 1-morphisms satisfy the group 
axioms strictly, as equations.  However, there is also an extensive 
theory of `weak' 2-groups, where the 1-morphisms obey the group axioms 
only up to invertible 2-morphisms \cite{BaezLauda:2groups}.  
Following this line of thought, we may also define weak Lie 2-algebras
\cite{Roytenberg}, and the Lie 2-algebra $\sstring_k(\g)$ is one of 
these where only the Jacobi identity fails to hold strictly.  

The beauty of weak Lie 2-algebras is that $\sstring_k(\g)$ is
very easy to describe in these terms.  In particular, it is
finite-dimensional.  The hard part is constructing
a weak Lie 2-group corresponding to this weak Lie 2-algebra.  It is
easy to check that any strict Lie 2-algebra has a corresponding strict
Lie 2-group.  Weak Lie 2-algebras are more tricky
\cite{BaezLauda:2groups}.  Baez, Crans, Schreiber and Stevenson
\cite{BCSS} dodged this problem by showing that the string Lie 
2-algebra is \emph{equivalent} (in some precise sense) to a strict 
Lie 2-algebra, which however is infinite-dimensional.  They then 
constructed the infinite-dimensional strict Lie 2-group corresponding 
to this strict Lie 2-algebra.   This is just $\STRING_k(G)$ as described 
above.

On the other hand, a finite-dimensional model of the string 2-group 
was recently introduced by Schommer-Pries \cite{Schommer-Pries}.  
This uses an improved definition of `weak Lie 2-group', based on
an important realization: the correct maps between smooth groupoids 
are not the smooth functors, but something more general \cite{Lerman}.  
We have already mentioned this in our discussion of connection:
a smooth functor $\hol \maps \P_1(M) \to G$ is a connection on the
trivial principal $G$-bundle over $M$, while one of these more general 
maps is a connection on an \emph{arbitrary} principal $G$-bundle over
$M$.  If we take this lesson to heart, we are led into the world of 
`stacks'---and in that world, we can find a finite-dimensional version 
of the string 2-group.

There has also been progress on constructing weak Lie $n$-groups from
weak Lie $n$-algebras for $n > 2$.  Getzler \cite{Getzler} and
Henriques \cite{Henriques} have developed an approach that works for
all $n$, even $n = \infty$.  Their approach is able to handle weak Lie
$\infty$-algebras of a sort known as `$L_\infty$-algebras'.  Quite
roughly, the idea is that in an $L_\infty$-algebra, the Jacobi
identity holds only weakly, while the antisymmetry of the bracket
still holds strictly.

In fact, $L_\infty$-algebras were developed by Stasheff and
collaborators \cite{MSS,SS} before higher gauge theory became
recognized as a subject of study.  But more recently, Sati, Schreiber,
Stasheff \cite{Sati,SSS} have developed a lot of higher gauge theory
with the help of $L_\infty$-algebras. Thanks to their work, it is
becoming clear that superstring theory, supergravity and even the
mysterious `$M$-theory' have strong ties to higher gauge theory.  For
example, they argue that 11-dimensional supergravity can be seen as a
higher gauge theory governed by a certain `Lie 3-superalgebra' which
they call $\sugra(10,1)$.  The number 3 here relates to the 2-brane
solutions of 11-dimensional supergravity: just as parallel transport
of strings is described by 2-connections, the parallel transport of
2-branes is described by 3-connections, which in the supersymmetric
case involve Lie 3-superalgebras.

In fact, $\sugra(10,1)$ is one of a family of four Lie 3-superalgebras
that extend the Poincar\'e Lie superalgebra in dimensions 4, 5, 7 and
11.  These can be built via a systematic construction starting from
the four normed division algebras: the real numbers, the complex
numbers, the quaternions and the octonions \cite{BaezHuerta:susy2}.
These four algebras also give rise to Lie 2-superalgebras extending
the Poincar\'e Lie superalgebra in dimensions 3, 4, 6, and 10.  The
Lie 2-superalgebras are related to superstring theories in dimensions
3, 4, 6, and 10, while the Lie 3-superalgebras are related to
super-2-brane theories in dimensions 4, 5, 7 and 11.  All these
theories, and even their relation to division algebras, have been
known since the late 1980s \cite{Duff}.  Higher gauge theory provides
new insights into the geometry of these theories.  In particular, the
work of D'Auria, Castellani and Fr\'e \cite{TheCube} can be seen as
\emph{implicitly} making extensive use of Lie $n$-superalgebras---but
this only became clear later, through the work of Sati, Schreiber 
and Stasheff \cite{SSS}.

Alas, explaining these fascinating issues in detail would vastly
expand the scope of this paper.  We should instead return to simpler
things: gauge transformations, curvature, and nontrivial 2-bundles.

\section{Further Topics}
\label{last}

So far our introduction to higher gauge theory has neglected the most
important topic of all: \emph{gauge transformations!}  We have also
said nothing about curvature or nontrivial 2-bundles.  Now it is time
to begin correcting these oversights.  

\subsection{Gauge Transformations}
\label{gauge}

First consider ordinary gauge theory.  Suppose that $M$ is a manifold
and $G$ is a Lie group.  Then a \define{gauge transformation} on the
trivial principal $G$-bundle over $M$ simply amounts to a smooth
function
\[              g \maps M \to G , \]
while a connection on this bundle can be seen as a $\g$-valued 1-form.  
A gauge transformation $g$ acts on a connection $A$ to give a new connection
$A'$ as follows:
\[            A' = g A g^{-1} + g\, dg^{-1}   .\]
This formula makes literal sense if $G$ is a group of matrices: then
$\g$ also consists of matrices, so we can freely multiply
elements of $G$ with elements of $\g$.  If $G$ is an arbitrary 
Lie group the formula requires a bit more careful interpretation, but
it still makes sense.   A well-known calculation says the curvature
$F' = dA' + A' \wedge A'$ of the gauge-transformed connection is
just the curvature of the original connection conjugated by $g$:
\[          F' = g F g^{-1}  .\]

In higher gauge theory the formulas are similar, but a bit more
complicated.  Suppose $M$ is a manifold and $\G$ is a Lie 2-group with
crossed module $(G,H,t,\alpha)$.  It will be helpful to take
everything in this crossed module and differentiate it.  Doing this,
we get:
\begin{itemize}
\item
the Lie algebra $\g$ of $G$,
\item
the Lie algebra $\h$ of $H$,
\item 
the Lie algebra homomorphism $\dt \maps \h \to \g$ obtained by
differentiating $t \maps H \to G$, and
\item 
the Lie algebra homomorphism $\dalpha \maps \g \to \aut(H)$ obtained
by differentiating $\alpha \maps G \to \Aut(H)$.
\end{itemize}
Here $\aut(H)$ is the Lie algebra of $\Aut(H)$.  It is best to think 
of this as the Lie algebra of \define{derivations} of $\h$: that is, linear
maps $D \maps \h \to \h$ such that
\[          D[x,y] = [Dx,y] + [x,Dy] . \]
If we differentiate the two equations in the definition of a
crossed module, we obtain the \define{\boldmath $\g$-equivariance 
of $\dt$}:
\[      \dt(\dalpha(x)(y)) = [x,\dt(y)]  \]
and the \define{infinitesimal Peiffer identity}:
\[      \dalpha(\dt(y))(y') = [y,y']  \]
where $x \in \g$ and $y,y' \in \h$.  In case the reader is curious: we
write $\dt$ and $\dalpha$ instead of $dt$ and $d\alpha$ because later
we will do computations involving these maps and also differential
forms, where $d$ stands for the exterior derivatve.

A quadruple $(\g, \h, \dt, \dalpha)$ of Lie algebras and homomorphisms
obeying these two equations is called an \define{infinitesimal crossed
module}.  Just as crossed modules are a way of working with 2-groups,
infinitesimal crossed modules are a way of working with Lie 2-algebras
\cite{BaezCrans}.  Any infinitesimal crossed module comes from a Lie
2-group, and this Lie 2-group is unique if we demand that $G$ and $H$
be connected and simply connected.

But we digress!  We have introduced infinitesimal crossed modules in order
to say how gauge transformations act on 2-connections.  A {\bf gauge
transformation} of the trivial $\G$-2-bundle over $M$ consists of two
pieces of data:
\begin{itemize}
\item
a smooth function $g \maps M \to G$, 
\item
an $\h$-valued 1-form $a$ on $M$.
\end{itemize}
Why \emph{two} pieces of data?  Perhaps this should not be so
surprising.  Remember, a 2-connection also consists of two pieces of
data:
\begin{itemize}
\item
a $\g$-valued 1-form $A$ on $M$, 
\item
an $\h$-valued 2-form $B$ on $M$ satisfying $\dt(B) = F$.
\end{itemize}

Breen and Messing \cite{BreenMessing} worked out how gauge
transformations act on connections on nonabelian gerbes, and their
work was later generalized to 2-connections on arbitrary principal
2-bundles \cite{BaezSchreiber, SchreiberWaldorf:gerbes}.  Here we
merely present the formulas.  A gauge transformation $(g,a)$ acts on a
2-connection $(A,B)$ to give a new 2-connection $(A',B')$ as follows:
\[       
\begin{array}{ccl}
A' &=& g A g^{-1} \; + \; g\, dg^{-1} \; + \; \dt(a)  \\  \\
B' &=& \alpha(g)(B) \; + \; \dalpha(A') \wedge a \; + \; da \; - \; 
a \wedge a
\end{array}
\]
The second formula requires a bit of explanation.  In the first term
we compose $\alpha \maps G \to \Aut(H)$ with $g \maps M \to
G$ and obtain an $\Aut(H)$-valued function $\alpha(g)$, which then
acts on the $\h$-valued 2-form $B$ to give a new $\h$-valued 2-form
$\alpha(g)(B)$.  In the second term we start by composing $A'$ with
$\dalpha$ to obtain an $\aut(H)$-valued 1-form $\dalpha(A')$.  Then
we wedge this with $a$, letting $\aut(H)$ act on $\h$ as part of this 
process, and obtain an $\h$-valued 2-form.

As a kind of consistency check and test of our understanding, let us
see why the gauge-transformed 2-connection $(A',B')$ satisfies the
equation $\dt(B') = F'$.  First let us compute the curvature 2-form
$F'$ of the gauge-transformed 2-connection:
\[
\begin{array}{ccl}
F' &=& dA' + A' \wedge A'  \\
   &=& d(g A g^{-1} + g\, dg^{-1} + \dt(a)) \; + \\
   &&    (g A g^{-1} + g\, dg^{-1} + \dt(a)) \wedge
       (g A g^{-1} + g\, dg^{-1} + \dt(a)) 
\end{array}
\]
This looks like a mess---but except for the terms containing $\dt(a)$,
this is just the usual mess we get in ordinary gauge theory when
we compute the curvature of a gauge transformed connection.  So, we
have:
\[
\begin{array}{ccl}
F'    &=& gFg^{-1}\; + \; d(\dt(a)) \; + \; \dt(a) \wedge A' 
\; + \; A' \wedge \dt(a) - \dt(a) \wedge \dt(a)  \\
&=& gFg^{-1}\; + \; \dt(da) \; + \; [\dt(a), A'] - \dt(a) \wedge \dt(a)      
\end{array}
\]
where we use the fact that $d(\dt(a)) = \dt(da)$ and rewrite
$A' \wedge \dt(a) + \dt(a) \wedge A'$ as a graded commutator.

On the other hand, we have
\[
\begin{array}{ccl}
\dt(B') &=& \dt(\alpha(g)(B)) \; + \; \dt(\dalpha(A') \wedge a) \; + \;
\dt(da - a \wedge a) \qquad \qquad
\end{array}
\]
The $G$-equivariance of $t$ implies that
$\dt(\alpha(g)(B)) = g\dt(B)g^{-1}$, and
the $\g$-equivariance of $\dt$ implies that 
$\dt(\dalpha(A') \wedge a) = [A', \, \dt(a)]$.  So, we see that
\[
\begin{array}{ccl}
\dt(B') &=& g\dt(B)g^{-1} \; + \; [A',\, \dt(a)] \; + \;
\dt(da - a \wedge a) \qquad \qquad
\end{array}
\]
and thus $\dt(B') = F'$, as desired.

\subsection{Curvature}
\label{curvature}

Suppose $\G$ is a Lie 2-group whose crossed module is $(G,H,t,
\alpha)$, and let $(\g,\h,\dt,\dalpha)$ be the corresponding
differential crossed module.  Suppose we have a \define{connection} on
the trivial $\G$-2-bundle over $M$: that is, a $\g$-valued 1-form $A$
and an $\h$-valued 2-form $B$.  

As in ordinary gauge theory, we may define the \define{curvature} of
this connection to be the $\g$-valued 2-form given by:
\[                 F = dA + A \wedge A  .\]
We also have another $\g$-valued 2-form, the \define{fake curvature}
$F - \dt(B)$.  Recall from Section \ref{2-connections} that only a
connection with vanishing fake curvature counts as a 2-connection.  In
other words, we need $\dt(B) = F$ to obtain well-defined parallel
transport over surfaces.  

We may also define the \define{2-curvature} of a connection in higher gauge
theory.  This is the $\h$-valued 3-form given by:
\[                 Z = dB + \dalpha(A) \wedge B  .\]
In the second term here, we compose $\dalpha \maps \g \to \aut(H)$
with the $\g$-valued 1-form $A$ and obtain an $\aut(H)$-valued
function $\dalpha(g)$.  Then we wedge this with $B$, letting $\aut(H)$
act on $\h$ as part of this process, and obtain an $\h$-valued 2-form.

The intuitive idea of 2-curvature is this: just as the curvature 
describes the holonomy of a connection around an infinitesimal loop, 
the 2-curvature describes the holonomy of a 2-connection over an 
infinitesimal 2-sphere.  This can be made precise using formulas for
holonomies over surfaces \cite{MartinsPicken:2007, SchreiberWaldorf:gerbes}

If the 2-curvature of a 2-connection vanishes, the holonomy over a surface
will not change if we apply a smooth homotopy to that surface while
keeping its edges fixed.  A 2-connection whose curvature and 2-curvature
both vanish truly deserves to be called \define{flat}.  We have seen
flat 2-connections already in our discussion of 4-dimensional $BF$
theory in Section \ref{tangent}: the solutions of this theory are 
flat 2-connections.

\subsection{Nontrivial 2-Bundles}
\label{nontrivial}

So far we have implicitly been looking at 2-connections on trivial
2-bundles.  This is fine locally.  But there are also interesting
issues involving nontrivial 2-bundles, which become crucial when we
work globally.   

A careful treatment of 2-bundles would require some work, and the
reader interested in this topic would do well to start with Moerdijk's
introductory paper on `stacks' and `gerbes' \cite{Moerdijk}.  Here we
take a less sophisticated approach: we simply describe how to build a
principal 2-bundle and put a 2-connection on it.  Since we do not say
when two principal 2-bundles built this way are the `same', our treatment
is incomplete.  The reader can find more details elsewhere
\cite{BaezStevenson, BJ, Bartels:2004, Breen:2006, Breen:2008,
BreenMessing, SchreiberWaldorf:gerbes}.  We warn the reader that
almost every paper in the literature uses different notation, sign
conventions, and so forth.  
 
First recall ordinary gauge theory: suppose $G$ is a Lie group and $M$
a manifold.  In this case we can build a principal $G$-bundle over $M$
using transition functions.  First, write $M$ as the union of open
sets or \define{patches} $U_i \subseteq M$:
\[ M = \bigcup_i U_i. \]
Then, choose a smooth \define{transition
function} on each double intersection of patches:
\[ g_{ij} \maps U_i \cap U_j \to G .\]
These transition functions give gauge transformations.  We can build a
principal $G$-bundle over all of $M$ by gluing together trivial bundles
over the patches with the help of these gauge transformations.
However, this procedure will only succeed if the transition functions
satisfy a consistency condition on each triple intersection:
\[ g_{ij}(x) g_{jk}(x) = g_{ik}(x) \]
for all $x \in U_i \cap U_j \cap U_k$.  This equation is called a
\define{cocycle condition}.  We can visualize it as a triangle:
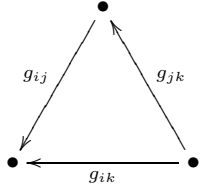
\[
\xy 
(0,0)*+{\bullet}="1"; 
(-12,-20.78)*+{\bullet}="2"; 
(12,-20.78)*+{\bullet}="3"; 
{\ar^-{g_{ik}} "3";"2"}; 
{\ar_-{g_{jk}} "3";"1"}; 
{\ar_-{g_{ij}} "1";"2"}; 
\endxy
\]
where we suppress the variable $x$ for the sake of readability.
The idea is that this triangle should `commute': the direct way
of identifying points in the trivial bundle in the $i$th patch 
to points in the trivial bundle over the $k$th patch should match 
the indirect way which proceeds via the $j$th patch.

A similar but more elaborate recipe works for higher gauge theory.
Now let $\G$ be a Lie 2-group with crossed module
$(G, H, t, \alpha)$.  To build a $\G$-2-bundle, we start by
choosing transition functions on double intersections of patches:
\[ g_{ij} \maps U_i \cap U_j \to G \]
However, now it makes sense to replace the equation in
the cocycle condition by a 2-morphism!  So, for each triple 
intersection we choose 2-morphisms
in $\G$:
\[ \gamma_{ijk}(x) \maps g_{ij}(x) \, g_{jk}(x) \Rightarrow g_{ik}(x) \]
depending smoothly on $x \in U_i \cap U_j \cap U_k$.   We can again
visualize these as triangles:
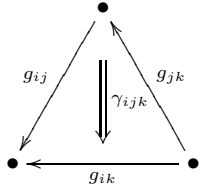
\[
\xy 
(0,0)*+{\bullet}="1"; 
(-12,-20.78)*+{\bullet}="2"; 
(12,-20.78)*+{\bullet}="3"; 
{\ar^-{g_{ik}} "3";"2"}; 
{\ar_-{g_{jk}} "3";"1"}; 
{\ar_-{g_{ij}} "1";"2"}; 
{\ar@2{->}^-{\gamma_{ijk}} (0,-7)*{};(0,-18)*{}};
\endxy
\]
But now we demand that these 2-morphisms themselves obey
a cocycle condition on quadruple intersections of patches.  
As we ascend the ladder of higher gauge theory, triangles become
tetrahedra and then higher-dimensional simplexes.  In this case, the
cocycle condition says that this tetrahedron commutes:
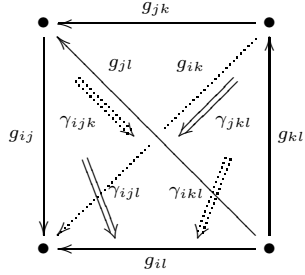
\[
{\xy 
(15,-15)*+{\bullet}="1"; 
(15,15)*+{\bullet}="2"; 
(-15,15)*+{\bullet}="3"; 
(-15,-15)*+{\bullet}="4"; 
(0,0)*+{}="0"; 
{\ar_-{g_{kl}} "1";"2"}; 
{\ar_-{g_{jk}} "2";"3"}; 
{\ar_-{g_{ij}} "3";"4"};
{\ar^-{g_{il}} "1";"4"};
{\ar@{.}_(0.5){\tiny{g_{ik}}} "2";"0"};
{\ar@{.>} "0";"4"};
{\ar_(0.75){g_{jl}} "1";"3"};
{\ar@2{->}^-{\gamma_{ijl}} (-9.5,-3)*{};(-5.5,-13.5)*{}};
{\ar@2{->}^-{\gamma_{jkl}} (10.5,7.5)*{};(3,0)*{}};
{\ar@2{.>}_-{\tiny{\gamma_{ijk}}} (-10.5,7.5)*{};(-3,0)*{}};
{\ar@2{.>}_-{\tiny{\gamma_{ikl}}} (9.5,-3)*{};(5.5,-13.5)*{}};
\endxy}
\]

\noindent
By saying that this tetrahedron `commutes', we mean that the composite
of the front two sides equals the composite of the back two sides:
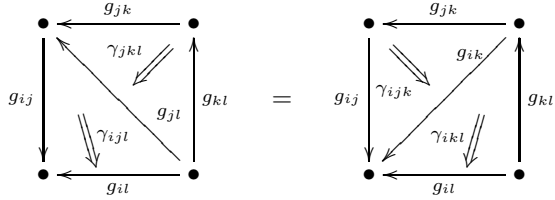
\[
{\xy 
(10,-10)*+{\bullet}="1"; 
(10,10)*+{\bullet}="2"; 
(-10,10)*+{\bullet}="3"; 
(-10,-10)*+{\bullet}="4"; 
{\ar_-{g_{kl}} "1";"2"}; 
{\ar_-{g_{jk}} "2";"3"}; 
{\ar_-{g_{ij}} "3";"4"};
{\ar^-{g_{il}} "1";"4"};
{\ar_(0.3){g_{jl}} "1";"3"};
{\ar@2{->}^-{\gamma_{ijl}} (-5,-2)*{};(-3,-9)*{}};
{\ar@2{->}_-{\gamma_{jkl}} (7,7)*{};(2,2)*{}};
\endxy}
\quad = \quad
{\xy 
(10,-10)*+{\bullet}="1"; 
(10,10)*+{\bullet}="2"; 
(-10,10)*+{\bullet}="3"; 
(-10,-10)*+{\bullet}="4"; 
{\ar_-{g_{kl}} "1";"2"}; 
{\ar_-{g_{jk}} "2";"3"}; 
{\ar_-{g_{ij}} "3";"4"};
{\ar^-{g_{il}} "1";"4"};
{\ar_{} "2";"4"}; 
(3.5,6)*+{\scriptstyle{g_{ik}}};
{\ar@2{->}_(0.8){\gamma_{ijk}} (-7,7)*{};(-2,2)*{}};
{\ar@2{->}_-{\gamma_{ikl}} (5,-2)*{};(3,-9)*{}};
\endxy}
\]
We need whiskering to compose the 2-morphisms in this diagram, as explained
near the end of Section \ref{2-connections}.  So in equations,
the tetrahedral cocycle condition says that:
\[
\gamma_{ijl} \cdot (g_{ij} \circ \gamma_{jkl}) = 
 \gamma_{ikl} \cdot (\gamma_{ijk} \circ g_{kl}) 
\]
where $\cdot$ stands for vertical composition and $\circ$ stands
for whiskering.

We can describe this cocycle condition in a more down-to-earth
manner if we use Theorem \ref{crossed_module}, which says that a
2-morphism
\[ \gamma_{ijk} \maps g_{ij} \, g_{jk} \Rightarrow g_{ik} \]
is the same as an element $h_{ijk} \in H$ such that
\[   
t(h_{ijk}) \, g_{ij} \, g_{jk} = g_{ik} .
\]
This theorem also gives formulas for vertical and horizontal 
composition in terms of the groups $G$ and $H$.  Since whiskering by 
a morphism is horizontal composition with its identity 2-morphism, 
we can also express whiskering in these terms.  So,
a little calculation---a wonderful exercise for the 
would-be higher gauge theorist---shows that:
\[
h_{ijl} \; \, \alpha(g_{ij})(h_{jkl}) = h_{ikl} \; h_{ijk} 
\]
where $\alpha$ is the action of $G$ on $H$.  

There is no need to have $g_{ii} = 1$ in this formalism; we should
really choose a 2-morphism from $g_{ii}$ to $1$.  However, without loss 
of generality, we can assume that $g_{ii} = 1$ and set this 2-morphism 
equal to the identity.  We can also assume that $h_{ijk} = 1$ whenever 
two or more of the indices $i,j,k$ are equal.  The reason is that
Bartels \cite{Bartels:2004} has shown that any principal 2-bundle is
equivalent to one for which these simplifying assumptions hold.  
We will make these assumptions in what follows.  For the full story 
without these assumptions, see Schreiber and Waldorf 
\cite{SchreiberWaldorf:gerbes}.

Now consider connections.  Again, it helps to begin by reviewing
the story for ordinary gauge theory.  Suppose we have a manifold $M$
written as a union of patches $U_i$, and suppose we have principal
$G$-bundle over $M$ built using transition functions $g_{ij}$.  To put
a connection on this bundle, we first put a connection on the trivial
bundle over each patch: that is, for each $i$, we choose a $\g$-valued
1-form $A_i$ on $U_i$.  But then we must check to see if these fit
together to give a well-defined connection on all of $M$.  For this,
we need the gauge transform of the connection on the $j$th patch to
equal the connection on the $i$th patch:
\[         A_i = g_{ij} A_j g_{ij}^{-1} + g_{ij} dg_{ij}^{-1}  \]
on each double intersection $U_i \cap U_j$.   

The story is similar for 2-connections.  Suppose we have a principal
$\G$-2-bundle over $M$ built using transition functions $g_{ij}$ and
$h_{ijk}$ as described above.  To equip this 2-bundle with a
2-connection, we first put a 2-connection on the trivial 2-bundle over
each patch.  So, on each open set $U_i$ we choose a $\g$-valued 1-form
$A_i$ and an $\h$-valued 2-form $B_i$ with $\dt(B_i) = F_i$.  But then
we must fit these together to get a 2-connection on all of $M$.

For this, we should follow the ideas from Section \ref{gauge} on how
gauge transformations work in higher gauge theory.  So, we choose
an $\h$-valued 1-form $a_{ij}$ on each double intersection $U_i \cap
U_j$, and require that
\[       
\begin{array}{ccl}
A_i &=& g_{ij} A_j g_{ij}^{-1} \, + \, g_{ij}\, dg_{ij}^{-1} \, + 
\, \dt(a_{ij})  \\  \\
B_i &=& \alpha(g_{ij})(B_j) \, + \, \dalpha(A_i) \wedge a_{ij} \, + \, 
da_{ij} \, - \, a_{ij} \wedge a_{ij} .
\end{array}
\]
These equations say that the 2-connection $(A_i,B_i)$ is a
gauge-transformed version of $(A_j,B_j)$.  The appearance of $A_i$ on
the right-hand side of the second equation is not a typo!  Finally,
the 1-forms $a_{ij}$ must obey a consistency condition on triple
intersections:
\begin{equation}
\label{compatibility}
  h_{ijk}^{-1} \, \dalpha(A_i)(h_{ijk}) \, + \,
   h_{ijk}^{-1} \, dh_{ijk}    \, + \, 
   \alpha(g_{ij})(a_{jk}) \, + \, a_{ij} = 
   h_{ijk}^{-1}\, a_{ik}\, h_{ijk}  
\end{equation}
Where does this consistency condition come from?  Indeed, what does
it even mean?  We have not yet defined `$h^{-1} \dalpha(A)(h)$' when 
$A \in \g$ and $h$ is an element of the group $H$.  

We could systematically derive this condition from a more conceptual
approach to 2-connections \cite{BaezSchreiber,
SchreiberWaldorf:gerbes}, but it will be marginally less stressful to
motivate it as follows.  For every triple intersection $U_i \cap U_j
\cap U_k$ we have three equations relating $A_i, A_j$ and $A_k$:
\[       
\begin{array}{ccl}
A_i &=& g_{ij} A_j g_{ij}^{-1} \, + \, g_{ij}\, dg_{ij}^{-1} \, + 
\, \dt(a_{ij}) \\   \\
A_j &=& g_{jk} A_k g_{jk}^{-1} \, + \, g_{jk}\, dg_{jk}^{-1} \, + 
\, \dt(a_{jk}) \\   \\
A_k &=& g_{ki} A_i g_{ki}^{-1} \, + \, g_{ki}\, dg_{ki}^{-1} \, + 
\, \dt(a_{ki}) 
\end{array}
\]
The first equation expresses $A_i$ in terms of $A_j$.
We can substitute the second equation in the first to get a formula for 
$A_i$ in terms of $A_k$.  Then we can use the third equation
to get a formula for $A_i$ in terms of itself!  We would like to
be able to simplify this formula to get simply $A_i = A_i$.  
The consistency condition, Equation (\ref{compatibility}), ensures that
we can do this.

This calculation is a bit of a workout; let us see how it goes.
We begin by doing the substitutions:
\[
\begin{array}{ccl}
A_i &=& g_{ij} A_j g_{ij}^{-1} \, + \, g_{ij}\, dg_{ij}^{-1} \, + 
\, \dt(a_{ij}) \\   \\
    &=& g_{ij} \left(g_{jk} A_k g_{jk}^{-1} \, + \, g_{jk}\, dg_{jk}^{-1} \, + 
\, \dt(a_{jk}) \right) g_{ij}^{-1} \, + \, g_{ij}\, dg_{ij}^{-1} \, + 
\, \dt(a_{ij}) \\   \\
    &=& g_{ij} \left(g_{jk} \left( g_{ki} A_i g_{ki}^{-1} + 
g_{ki} dg_{ki}^{-1}  + \dt(a_{ki})  \right) g_{jk}^{-1} + 
g_{jk}\, dg_{jk}^{-1} + \dt(a_{jk}) \right) g_{ij}^{-1}  + \\  
&& g_{ij}\, dg_{ij}^{-1} \, + \, \dt(a_{ij}) .
\end{array}
\]
Then we do a bit of simplification:
\[  
\begin{array}{ccl}
A_i &=& g_{ij} g_{jk} g_{ki} A_i (g_{ij} g_{jk} g_{ki})^{-1} \, + \,
g_{ij} g_{jk} g_{ki} d(g_{ij} g_{jk} g_{ki})^{-1} \, +  \\
&& g_{ij} g_{jk}  \dt(a_{ki})  (g_{ij} g_{jk})^{-1} \, + \,
g_{ij} \dt(a_{jk}) g_{ij}^{-1} \, + \, \dt(a_{ij}) .
\end{array}
\]
Since $t(h_{ijk}) \, g_{ij} \, g_{jk} = g_{ik}$ and by our assumptions
 $g_{ik}^{-1} = g_{ki}$, we have
\[   g_{ij} g_{jk} g_{ki} = t(h_{ijk})^{-1}   \]
so
\begin{equation}
\label{step1}
\begin{array}{ccl}
0 &=& t(h_{ijk})^{-1} \, A_i \, t(h_{ijk}) - \, A_i \, + \,  
         t(h_{ijk})^{-1} \, dt(h_{ijk})    \, +  \\  
&& t(h_{ijk})^{-1} g_{ik} \dt(a_{ki}) g_{ik}^{-1} t(h_{ijk})  \, + \,
g_{ij} \dt(a_{jk}) g_{ij}^{-1} \, + \, \dt(a_{ij}) .
\end{array}
\end{equation}

Now, if $G$ is a matrix group, we can freely multiply group elements
and Lie algebra elements.  Then for any $h \in H$ and $A \in \g$ we
have
\begin{equation}
\label{step2}
   t(h)^{-1} A t(h) \, - \, A =  t(h)^{-1} [A, t(h) ] .
\end{equation}
This will allow us to simplify the first two terms in Equation (\ref{step1}).
Moreover, for any $g \in G$, $h \in H$ we have
\[    t(h^{-1} \alpha(g) h) = t(h)^{-1} t(\alpha(g)(h)) = 
      t(h)^{-1} g t(h) g^{-1}  .\]
Taking $g = \exp(s A)$ for $A \in \g$ and differentiating 
this equation with respect to $s$ at $s = 0$, we get
\begin{equation}
\label{step3}
\dt(h^{-1} \dalpha(A)(h)) =  t(h)^{-1} A t(h) \, - \, A 
\end{equation}
where on the left side $\dalpha$ means the derivative of the map
$\alpha \maps G \times H \to H$ with respect to its first argument,
while $\dt$ is the derivative of $t$.  Here are we extending our
previous definitions of $\dalpha$ and $\dt$.  
Combining Equations (\ref{step2}) and (\ref{step3}) we see that
\[       t(h)^{-1} [A, t(h)] =  \dt\left(h^{-1} \dalpha(A)(h)\right)  .\]
In fact this result holds even when $G$ is a non-matrix Lie group, 
as long as we carefully make sense of both sides.

Using this result, we can rewrite Equation (\ref{step1}) as follows:
\[  
\begin{array}{ccl}
0 &=&    \dt(h_{ijk}^{-1} \, \dalpha(A_i)(h_{ijk})) \, + \,
   t(h_{ijk})^{-1} \, dt(h_{ijk})    \, +  \\  
&& t(h_{ijk})^{-1} g_{ik} \, \dt(a_{ki})
   \, g_{ik}^{-1} t(h_{ijk}) \, + \, g_{ij} \, \dt(a_{jk}) \, g_{ij}^{-1} \,
   + \, \dt(a_{ij})
\end{array}
\]
Each term in the above equation is $\dt$ applied to an $\h$-valued 1-form. 
Writing down all these $\h$-valued 1-forms, we see that the above equation 
will be true if this condition holds:
\[  
0 =  h_{ijk}^{-1} \dalpha(A_i) (h_{ijk}) \, + \,
   h_{ijk}^{-1} dh_{ijk}    \, + \,   \\  \\
  h_{ijk}^{-1} \alpha(g_{ik})(a_{ki}) h_{ijk}  \, + \,  
   \alpha(g_{ij})(a_{jk}) \, + \, a_{ij} 
\]
This is our consistency condition in disguise!  To remove the
disguise, let us simplify it a bit further.  When $i = j$ this 
condition reduces to 
\[  
 a_{ik}  \, + \, \alpha(g_{ik})(a_{ki}) = 0.
\]
Reinserting this result we obtain
\[  
\begin{array}{ccl}
0 &=&  h_{ijk}^{-1} \, \dalpha(A_i) (h_{ijk}) \, + \,
   h_{ijk}^{-1} \, dh_{ijk}    \, - \,  
   h_{ijk}^{-1} \, a_{ik} \, h_{ijk}  \, + \, 
   \alpha(g_{ij})(a_{jk}) \, + \, a_{ij}
\end{array}
\]
\emph{Voil\`a!}  This is clearly equivalent to the consistency condition
we stated in the first place, Equation (\ref{compatibility}):
\[  
  h_{ijk}^{-1} \, \dalpha(A_i) (h_{ijk}) \, + \,
   h_{ijk}^{-1} \, dh_{ijk}    \, + \, 
   \alpha(g_{ij})(a_{jk}) \, + \, a_{ij} = 
   h_{ijk}^{-1} \, a_{ik} \, h_{ijk}  
\]

Now let us consider some examples.  Recall from Section \ref{abelian} that
$\mathrm{b}\U(1)$ is 2-group with one morphism and $\U(1)$ as its
group of 2-morphisms.  A \define{\boldmath $\U(1)$ gerbe} is principal
$\mathrm{b}\U(1)$-2-bundle.  Let's look at principal $\U(1)$-bundles
and then $\U(1)$ gerbes, to get a feel for how they are similar and
how they differ.

To build a principal $\U(1)$-bundle with a connection on it, we
choose transition functions
\[         g_{ij} \maps U_i \cap U_j \to \U(1) \]
such that 
\[ g_{ij} g_{jk} = g_{ik} \]
on each triple intersection.  
To put a connection on this bundle, we then choose a 1-form $A_i$ on 
each patch such that
\[ A_i = A_j + g_{ij} dg^{-1}_{ij} \]
on each double intersection $U_i \cap U_j$.  The curvature of 
this connection is then 
\[ F_i = dA_i \]
on the $i$th patch.  Note that $F_i = F_j$ on $U_i \cap U_j$, 
so we get a well-defined curvature 2-form $F$ on all of $M$.

To build a $\U(1)$ gerbe, we choose transition functions 
$h_{ijk} \maps U_i \cap U_j \cap U_k \to \U(1)$ such that
\[  h_{jkl} h_{ijl} = h_{ikl} h_{ijk} \]
on each quadruple intersection.  Remember, the 2-group
$\mathrm{b}\U(1)$ has only one morphism, the identity, so the
transition functions $g_{ij}$ are trivial and can be ignored.  To put
a 2-connection on this gerbe, we must first choose a 2-form $B_i$ on
each patch.  Then we must choose a 1-forms $a_{ij}$ on each double
intersection.  We require that
\[ B_i = B_j + da_{ij}  \]
on each double intersection, and 
\[ a_{ij} + a_{jk} = a_{ik} + h_{ijk} \, dh^{-1}_{ijk} \]
on each triple intersection.   The \define{2-curvature} of this 
2-connection is then
\[ Z_i = dB_i \]
on the $i$th patch.  Note that $Z_i = Z_j$ on $U_i \cap U_j$, so we
get a well-defined 2-curvature 3-form $Z$ on all of $M$. 

There is a nice link between $\U(1)$ gerbes and cohomology, which
in fact is the reason they were invented in the first place.  
For any principal $\U(1)$-bundle with connection, the curvature $F$ is 
\define{integral}:
\[ \int_\Sigma F \in 2\pi \Z \]
for any closed surface $\Sigma$ mapped into $M$. In addition, 
$F$ is closed: 
\[ dF = 0. \] 
Conversely, any closed, integral 2-form $F$ on $M$ is the curvature of
some connection on a principal $\U(1)$-bundle over $M$.  Two different
connections on the same bundle have curvature 2-forms that differ by
an exact 2-form, so we get a well-defined element of the de Rham
cohomology $H^2(M,\R)$ from a principal $\U(1)$-bundle.  This idea can
be refined further, and the upshot is that principal $\U(1)$-bundles
over $M$ are classified by the cohomology group $H^2(M,\Z)$.

Similarly, for any $\U(1)$ gerbe, the curvature 3-form $Z$ is closed 
and integral, where the latter term now means that 
\[ \int_Y Z \in 2\pi \Z \]
for any closed 3-manifold $Y$ mapped into $M$.  Conversely, any such
3-form is the 2-curvature of some 2-connection on a $\U(1)$ gerbe over
$M$---and in fact, $\U(1)$ gerbes over $M$ are classified by
$H^3(M,\Z)$.  

This is just the beginning of a longer tale: namely, the story of
characteristic classes in higher gauge theory \cite{BaezStevenson,SSS}.
Indeed, though higher gauge theory is only in its infancy, there is
much more to say.   But our story ends here.  We invite the reader to 
go further.

\subsection*{Acknowledgements}

This paper is based on JH's notes of lectures given by JB 
at the 2nd School and Workshop on Quantum Gravity and Quantum
Geometry, held as part of the 2010 Corfu Summer Institute.  We thank
George Zoupanos, Harald Grosse, and everyone else involved with the
Corfu Summer Institute for making our stay a pleasant and productive
one.  We thank Urs Schreiber for many discussions of higher gauge
theory, and thank Tim van Beek, David Roberts and Elliot Schneider for catching 
some errors.  This research was partially supported by an FQXi grant.


\begin{thebibliography}{10}

\bibitem{AschieriCantiniJurco:2003}
P.~Aschieri, L.~Cantini, and B.~Jur{\v c}o, Nonabelian bundle gerbes,
their differential geometry and gauge theory,\
\textsl{Comm.\ Math.\ Phys.\ } \textbf{254} (2005), 367--400.  
Also available as \hepth{0312154}.

\bibitem{AschieriJurco:2004}
P.~Aschieri and B.~Jur{\v c}o, Gerbes, {M5}-brane anomalies and {$E_8$}
gauge theory, \textsl{JHEP} \textbf{10} (2004) 068.  Also available
as \hepth{0409200}.

\bibitem{AshtekarLewandowski:1994} 
A.\ Ashtekar and J.\ Lewandowski, Differential geometry on the 
space of connections via graphs and projective limits,
{\sl J.\ Geom.\ Phys.\ }{\bf 17} (1995), 191--230.
Also available as \hepth{9412073}.

\bibitem{Baez:BF}
J.\ Baez, Four-dimensional $BF$ theory as a topological quantum field 
theory, \textsl{Lett.\ Math.\ Phys.\ }{\bf 38} (1996), 129--143. 

\bibitem{Baez:spinfoam}
J.\ Baez, Spin foam models, \textsl{Class.\ Quantum Grav.\ }{\bf 15}
(1998), 1827--1858.  Also available as \grqc{9709052}.

J.\ Baez, An introduction to spin foam models of $BF$ theory and
quantum gravity, in \textsl{Geometry and Quantum Physics}, eds.\ H.\
Gausterer and H.\ Grosse, Springer, Berlin, 2000, pp.\ 25--93.

\bibitem{BBFW} 
J.\ Baez, A.\ Baratin, L.\ Freidel and D.\ Wise, Infinite-dimensional
representations of 2-groups.  Available as \arxiv{0812.4969}.

\bibitem{BaezCrans} 
J.\ Baez and A.\ Crans, Higher dimensional algebra VI: Lie 2-algebras,
\textsl{Theory and Applications of Categories} \textbf{12} (2004),
492--538.  Also available as \MMath{0307263}.

\bibitem{BCSS} 
J.\ Baez, A.\ Crans, U.\ Schreiber and D.\ Stevenson From loop groups
to 2-groups, \textsl{HHA} {\bf 9} (2007), 101--135.  Also available as
\Math{QA}{0504123}.

\bibitem{BaezCransWise} 
J.\ Baez, A.\ Crans and D.\ Wise, Exotic statistics for strings in 4d
$BF$ theory, \textsl{Adv.\ Theor.\ Math.\ Phys.\ }\textbf{11} (2007),
707--749.  Also available as \grqc{0603085}.


\bibitem{BaezHoffnungRogers} 
J.\ Baez, A.\ Hoffnung, C.\ Rogers, Categorified symplectic geometry
and the classical string, \textsl{Comm.\ Math.\ Phys.} {\bf 293}
(2010), 701--715.  Also available as \arxiv{0808.0246}.

\bibitem{BaezHuerta:susy2}
J.\ Baez and J.\ Huerta, Division algebras and supersymmetry II.
Available as \arxiv{1003.3436}.

\bibitem{BaezLauda:prehistory} J.\ Baez and A.\ Lauda, A
prehistory of $n$-categorical physics.  Available as
\arxiv{0908.2469}.

\bibitem{BaezLauda:2groups} J.\ Baez and A.\ Lauda, Higher
dimensional algebra V: 2-groups, \textsl{Theory and
Applications of Categories} \textbf{12} (2004), 423--491. Also
available as \MMath{0307200}.

\bibitem{BaezPerez} J.\ Baez and A.\ Perez, Quantization of strings
and branes coupled to $BF$ theory, \textsl{Adv.\ Theor.\ Math.\ 
Phys.\ }{\bf 11} (2007), 1--19.  Also available as \grqc{0605087}.

\bibitem{BaezRogers} J.\ Baez and C.\ Rogers, Categorified symplectic
geometry and the string Lie 2-algebra, to appear in \textsl{Homology,
Homotopy and Applications}.  Available as \arxiv{0901.4721}.

\bibitem{BaezSchreiber} J.\ Baez and U.\ Schreiber, Higher gauge
theory, in \textsl{Categories in Algebra, Geometry and
Mathematical Physics}, eds.\ A.\ Davydov et al, Contemp.\
Math.\ 431, AMS, Providence, 2007, pp.\ 7--30. Also available
\MMath{0511710}.

\bibitem{BaezSawin:1995}
J.\ Baez and S.\ Sawin, Functional integration on spaces of 
connections, \textsl{J.\ Funct.\ Analysis} {\bf 150} (1997), 1-26.
Also available as \qalg{9507023}.

\bibitem{BaezStevenson} J.\ Baez and D.\ Stevenson, 
The classifying space of a topological 2-group, in \textsl{Algebraic 
Topology: The Abel Symposium 2007}, eds.\ Nils Baas, Eric Friedlander, 
Bj\o rn Jahren and Paul Arne \O stv\ae r, Springer, Berlin, 2009.

\bibitem{BJ} 
I.\ Bakovic and B.\ Jur{\v c}o, The classifying topos of a topological
bicategory.  Available as \arxiv{0902.1750}.

\bibitem{Balachandran2}
A.\ P.\ Balachandran, F.\ Lizzi and G.\ Sparano, A new approach
to strings and superstrings, \textsl{Nucl.\ Phys.\ }{\bf B277} (1986),
359-387.

\bibitem{BaratinFreidel:3d} 
A.\ Baratin and L.\ Freidel, Hidden quantum gravity in 3d Feynman
diagrams, \textsl{Class.\ Quant.\ Grav.\ }{\bf 24} (2007), 1993--2026.
Also available as \grqc{0604016}. 

\bibitem{BaratinFreidel:4d}
A.\ Baratin and L.\ Freidel, Hidden quantum gravity in 4d Feynman
diagrams: emergence of spin foams, \textsl{Class.\ Quant.\ Grav.\
}{\bf 24} (2007), 2027--2060.  Also available as \hepth{0611042}.

\bibitem{BaratinWise}
A.\ Baratin and D.\ Wise, 2-Group representations for spin foams.
Available as \arxiv{0910.1542}.

\bibitem{BarrettMackaay} 
J.~W.~Barrett and M.~Mackaay, Categorical
representations of categorical groups, {\sl Th.\ Appl.\ Cat.} {\bf 16}
(2006), 529--557 Also available as \MMath{0407463}.

\bibitem{Bartels:2004}
T.~Bartels, Higher gauge theory: 2-bundles.  Available as
\hfill \break 
\Math{CT}{0410328}.

\bibitem{Benabou} Jean B{\'e}nabou, Introduction to bicategories, in
\textsl{Reports of the Midwest Category Seminar}, Springer, Berlin,
1967, pp.\ 1--77.

\bibitem{Borceux} F.\ Borceux, {\sl Handbook of Categorical Algebra 1:
Basic Category Theory}, Cambridge U.\ Press, Cambridge, 1994.

\bibitem{Breen:2006}
L.\ Breen, Notes on 1- and 2-gerbes, in \textsl{Towards Higher
Categories}, eds.\ J.\ Baez and P.\ May, Springer, Berlin, 2009,
pp.\ 193--235.  Also available as \MMath{0611317}.

\bibitem{Breen:2008}
L.~Breen, Differential geometry of gerbes and differential forms.
Available as \arxiv{0802.1833}.

\bibitem{BreenMessing} 
L.~Breen and W.~Messing, Differential geometry of gerbes,
\textsl{Adv.\ Math.\ }{\bf 198} (2005), 732--846.  Available as
\Math{AG}{0106083}.

\bibitem{Brylinski}
J.-L.~Brylinski, {\sl Loop Spaces, Characteristic Classes and Geometric
Quantization}, Birkh\"auser, Boston, 1993.

\bibitem{Brylinski:2000}
J.-L.~Brylinski, Differentiable cohomology of gauge groups.  Available as
\Math{DG}{0011069}.

\bibitem{BrylinskiMcLaughlin}
J.-L.~Brylinski and D.\ A.\ McLaughlin, The geometry of degree-four
characteristic classes and of line bundles on loop spaces I, 
\textsl{Duke Math.\ J.\ }{\bf 75} (1994), 603--638.  

J.-L.~Brylinski and D.\ A.\ McLaughlin, The geometry of degree-four
characteristic classes and of line bundles on loop spaces II, 
\textsl{Duke Math.\ J.\ }{\bf 83} (1996), 105--139.

\bibitem{CJM}
A.\ L.\ Carey, S.\ Johnson and M.\ K.\ Murray, Holonomy on D-branes,
\textsl{Jour.\ Geom.\ Phys.\ } {\bf 52} (2004), 186--216.  Also
available as \hepth{0204199}.

\bibitem{CKY}
L.~Crane, L.~Kauffman and D.~Yetter, State-sum
invariants of 4-manifolds I.  Available as \hepth{9409167}.

\bibitem{CraneSheppeard}
L.~Crane and M.~D.~Sheppeard, 2-Categorical Poincar\'e representations
and state sum applications, available as \MMath{0306440}.

\bibitem{CY}
L.\ Crane and D.\ Yetter, A categorical construction
of 4d TQFTs, in {\sl Quantum Topology,} eds.\ L.\ Kauffman and
R.\ Baadhio, World Scientific, Singapore, 1993, pp.\ 120-130.
Also available as \hepth{9301062}.

\bibitem{CraneYetter}
L.~Crane and D.~N.~Yetter, Measurable categories and 2-groups,
{\sl Appl.\ Cat.\ Str.} {\bf 13} (2005), 501--516.  Also available as
\MMath{0305176}.

\bibitem{DeDonder} 
T.\ DeDonder, \textit{Theorie Invariantive du Calcul des Variations},
Gauthier--Villars, Paris, 1935.

\bibitem{DMN} 
C.\ DeWitt-Morette, A.\ Maheshwari, and B.\ Nelson, Path integration
in phase space, \textit{Gen.\ Relativity Gravitation} \textbf{8}
(1977), 581--593.

\bibitem{TheCube}
L.\ Castellani, R.\ D'Auria and P.\ Fr\'e, \textsl{Supergravity 
and Superstrings: A Geometric Perspective}, World Scientific, 
Singapore, 1991.

\bibitem{Duff} M.\ J.\ Duff, Supermembranes: the first fifteen weeks, 
\textit{Class.\ Quantum Grav.\ }{\bf 5} (1988), 189--205.
Also available at 
\hfill \break 
\href{http://www-lib.kek.jp/cgi-bin/img_index?8708425}
{$\langle$http://www-lib.kek.jp/cgi-bin/img$\underline{\;}$index?8708425$\rangle$}.

\bibitem{EH} B.\ Eckmann and P.\ Hilton, Group-like structures in
categories, {\sl Math.\ Ann.\ }{\bf 145} (1962), 227-255.

\bibitem{EM}
S.\ Eilenberg and S.\ Mac Lane, General theory of
natural equivalences, \textsl{Trans.\ Amer.\ Math.\ Soc.}
\textbf{58} (1945), 231--294.

\bibitem{Ehresmann:1959}
C.~Ehresmann, Cat\'egories topologiques et cat\'egories differentiables,
{\sl Coll.\ G\'eom.\ Diff.\ Globale (Bruxelles, 1958)}, 
Centre Belge Rech.\ Math.\, Louvain, 1959, pp.\ 137--150.

\bibitem{Ehresmann:1963}
C.~Ehresmann, Cat\'egories structur\'ees, 
{\sl Ann.\ Ecole Norm.\ Sup.\ } \textbf{80} (1963), 349--426.

\bibitem{Ehresmann:1965}
C.~Ehresmann, {\sl Cat\'egories et Structures}, Dunod, Paris, 1965.

\bibitem{Elgueta2} J.~Elgueta, Representation theory of 2-groups on
Kapranov and Voevodsky 2-vector spaces, {\sl Adv.\ Math.\ }{\bf 213}
(2007), 53--92.  Also available as \MMath{0408120}.

\bibitem{Fairbairn}
W.\ Fairbairn and A.\ Perez, Extended matter coupled to $BF$ theory,
\textsl{Adv.\ Theor.\ Math.\ Phys.\ }{\bf D78}:024013 (2008).
Also available as \arxiv{0709.4235}.

\bibitem{FNS}
W.\ Fairbairn, K.\ Noui and F.\ Sardelli, Canonical analysis of
algebraic string actions.  Available as \arxiv{0908.0953}.

\bibitem{ForresterBarker} M.\ Forrester-Barker, Group objects and
internal categories, available as \Math{CT}{0212065}.

\bibitem{FreedWitten} D.\ S.\ Freed and E.\ Witten, Anomalies in
string theory with D-Branes, \S 6: Additional remarks, \textsl{Asian
J.\ Math.\ }{\bf 3} (1999) 819--852.  Also available as
\hepth{9907189}.

\bibitem{FreidelLouapre}
L.\ Freidel and D.\ Louapre: Ponzano--Regge model
revisited. I: Gauge fixing, observables and interacting spinning
particles, \textsl{Class.\ Quant.\ Grav.} \textbf{21} (2004), 5685--5726.
Also available as \hepth{0401076}.

L.\ Freidel and D.\ Louapre: Ponzano--Regge model
revisited. II: Equivalence with Chern--Simons.
Also available as \grqc{0410141}.

L.\ Freidel and E.\ Livine: Ponzano--Regge model
revisited. III: Feynman diagrams and effective field theory.
Also available as \hepth{0502106}.


\bibitem{Gawedski} 
K.\ Gawedzki, Topological actions in two-dimensional quantum field
theories, in \textsl{Nonperturbative Quantum Field Theory}, eds.\
G. t'Hooft, A.\ Jaffe, G.\ Mack P.\ K.\ Mitter and R.\ Stora, Plenum,
New York, 1988, pp.\ 101--141.

\bibitem{GawedskiReis}
K.\ Gawedzki and N.\ Reis, WZW branes and gerbes,
\textsl{Rev.\ Math.\ Phys.\ }{\bf 14} (2002), 1281--1334.
Also available as \hepth{0205233}.

\bibitem{Getzler}
E.\ Getzler, Lie theory for nilpotent $L_\infty$-algebras.
Available as \MMath{0404003}.

\bibitem{GIMM} M.\ Gotay, J.\ Isenberg, J.\ Marsden, and R.\ Montgomery,
Momentum maps and classical relativistic fields. Part I: covariant 
field theory, available as
\href{http://arxiv.org/abs/physics/9801019}{\texttt arXiv:physics/9801019}.

\bibitem{GirelliPfeiffer:2004}
F.~Girelli and H.~Pfeiffer,  Higher gauge theory - differential versus
integral formulation, \textsl{Jour.\ Math.\ Phys.\ } \textbf{45}
(2004), 3949--3971.  Also available as \hepth{0309173}.

\bibitem{GirelliPfeifferPopescu}
F.~Girelli, H.~Pfeiffer and E.\ M.\ Popescu, Topological higher gauge
theory - from $BF$ to $BFCG$ theory, \textsl{J.\ Math.\ Phys.\ }
{\bf 49}:032503, 2008.  Also available as \arxiv{0708.3051}.

\bibitem{GS} V.\ Guillemin and S.\ Sternberg, {\it Symplectic
Techniques in Physics}, Cambridge U.\ Press, Cambridge, 1984.


\bibitem{Henriques}
A.\ Henriques, Integrating $L_\infty$-algebras.  
Available as \MMath{0603563}.

\bibitem{Johnson} M.\ Johnson, The combinatorics of $n$-categorical pasting,
{\sl Jour.\ Pure Appl.\ Alg.\ }{\bf 62} (1989), 211--225.

\bibitem{Lack} S.\ Lack, A 2-categories companion, 
in \textsl{Towards Higher Categories}, eds.\ J.\ Baez and
P.\ May, Springer, 2009, pp.\ 105--191.  Also available as \MMath{0702535}.

\bibitem{Lerman} E.\ Lerman, Orbifolds as stacks?, available as 
\arxiv{0806.4160}.

\bibitem{LewandowskiThiemann:1999}
J.\ Lewandowski and T.\ Thiemann,
Diffeomorphism invariant quantum field theories of connections in
terms of webs, \textsl{ Class.\ Quant.\ Grav.\ }{\bf 16} (1999),
2299--2322.  Also available as \grqc{9901015}.

\bibitem{MW} S.\ Mac Lane and J.\ H.\ C.\ Whitehead, On the
3-type of a complex, {\sl Proc.\ Nat.\ Acad.\ Sci.\ }{\bf 36} (1950),
41--48.

\bibitem{MackaayPicken}
M.~Mackaay and R.~Picken, Holonomy and parallel transport for Abelian
gerbes, {\sl Adv.\ Math.\ } {\bf 170} (2002), 287--339.
Also available as \Math{DG}{0007053}.

\bibitem{MSS} 
M.\ Markl, S.\ Schnider and J.\ Stasheff,
{\sl Operads in Algebra, Topology and Physics},
AMS, Providence, Rhode Island, 2002.

\bibitem{MartinsPicken:2007} 
J.\ F.\ Martins and R.\ Picken, On two-dimensional holonomy.
Available as \arxiv{0710.4310}.

J.\ F.\ Martins and R.\ Picken, A cubical set approach to 2-bundles
with connection and Wilson surfaces.  Available as \arxiv{0808.3964}.

\bibitem{MartinsPicken:2009} 
J.\ F.\ Martins and R.\ Picken, The fundamental Gray 3-groupoid of a
smooth manifold and local 3-dimensional holonomy based on a 2-crossed
module.  Available as \arxiv{0907.2566}

\bibitem{Milnor} J.~Milnor, Remarks on infinite dimensional Lie groups,
{\sl Relativity, Groups and Topology II, (Les Houches, 1983)},
North-Holland, Amsterdam, 1984.

\bibitem{Moerdijk} 
I.\ Moerdijk, Introduction to the language of stacks and gerbes.
Available as \Math{AT}{0212266}.

\bibitem{MP} 
M.\ Montesinos and A.\ Perez, Two-dimensional topological field
theories coupled to four-dimensional $BF$ theory, \textsl{Phys.\ Rev.\ }
{\bf D77}:104020 (2008).  Also available as \arxiv{0709.4235}.

\bibitem{Murray} 
M.~K.~Murray, Bundle gerbes, {\sl J.\ London Math.\ Soc.\ }{\bf 54}
(1996), 403--416.  Also available as
\href{http://arxiv.org/abs/dg-ga/9407015}{arXiv:dg-ga/9407015}.

M.~K.~Murray, An introduction to bundle gerbes.  Available as
\arxiv{0712.1651}.

\bibitem{MurrayStevenson} M.~K.~Murray and D.\ Stevenson, Higgs
fields, bundle gerbes and string structures, \textit{Commun.\ Math.\
Phys.\ }{\bf 243} (2003), 541--555. Also available as \Math{DG}{0106179}.

\bibitem{Pfeiffer:2003}
H.~Pfeiffer, Higher gauge theory and a non-{A}belian generalization of
2-form electromagnetism, \textsl{Ann.\ Phys.\ } \textbf{308} (2003), 447--477.
Also available as \hepth{0304074}.

\bibitem{Power} 
A.\ J.\ Power, A 2-categorical pasting theorem, 
{\sl Jour.\ Alg.\ }{\bf 129} (1990), 439-445.

\bibitem{PressleySegal:1986} 
A.~Pressley and G.~Segal, {\sl Loop Groups}, Oxford U.\ Press, 
Oxford, 1986.

\bibitem{RobertsSchreiber} 
D.\ Roberts and U.\ Schreiber, The inner automorphism 3-group of a
strict 2-group.  Also available as \arxiv{0708.1741}.

\bibitem{Rovelli} C.\ Rovelli, \textsl{Quantum Gravity}, 
Cambridge U.\ Press, 2004.   Also available at
\href{http://www.cpt.univ-mrs.fr/~rovelli/book.pdf}
{$\langle$http://www.cpt.univ-mrs.fr/$\sim$rovelli/book.pdf$\rangle$}.

\bibitem{Roytenberg} D.\ Roytenberg, On weak Lie 2-algebras.
Available as \arxiv{0712.3461}.

\bibitem{Sati} H.\ Sati, Geometric and topological structures
related to M-branes.  Available as \arxiv{1001.5020}.

\bibitem{SSS} H.\ Sati, U.\ Schreiber and J.\ Stasheff,
$L_\infty$-algebras and applications to string- and
Chern--Simons $n$-transport.  Available as \arxiv{0801.3480}.

\bibitem{SS} M.\ Schlessinger and J.\ Stasheff, The Lie algebra
structure of tangent cohomology and deformation theory, {\sl Jour.\
Pure App.\ Alg.} \textbf{38} (1985), 313--322.

\bibitem{Schommer-Pries} C.\ Schommer-Pries, A finite-dimensional
string 2-group.  Available as \arxiv{0911.2483}.

\bibitem{Schreiber:cafe}
U.\ Schreiber, comments at the $n$-Category Caf\'e.
Available at 
\href{http://golem.ph.utexas.edu/category/2009/09/questions_on_ncurvature.html}{$\langle$http://golem.ph.utexas.edu/category/2009/09/questions$\underline{\;}$on$\underline{\;}$ncurvature.html$\rangle$}.

\bibitem{SchreiberWaldorf:functors} 
U.\ Schreiber and K.\ Waldorf, Parallel transport and functors,
\textsl{J.\ Homotopy Relat.\ Struct.\ }\textbf{4} (2009),
187--244. Also available as \arxiv{0705.0452}.

\bibitem{SchreiberWaldorf:gerbes} 
U.\ Schreiber and K.\ Waldorf, Smooth functors vs.\ differential
forms.  Available as \arxiv{0802.0663}.

U.\ Schreiber and K.\ Waldorf, Connections on non-abelian gerbes and
their holonomy.  Available as \arxiv{0808.1923}.


\bibitem{Stevenson}
D.~Stevenson, {\sl The Geometry of Bundle Gerbes}, 
Ph.D.\ thesis, University of Adelaide, 2000.  
Also available as \Math{DG}{0004117}.

\bibitem{StolzTeichner} 
S.\ Stolz and P.\ Teichner, What is an elliptic object?, in {\sl
Topology, Geometry and Quantum Field Theory: Proceedings of the 2002
Oxford Symposium in Honour of the 60th Birthday of Graeme Segal}, ed.\
U.\ Tillmann, Cambridge U.\ Press, Cambridge, 2004.

\bibitem{Waldorf} 
K.\ Waldorf, String connections and Chern--Simons
theory.  Available as \arxiv{0906.0117}.

\bibitem{Weyl} H.\ Weyl, Geodesic fields in the calculus of variation
for multiple integrals, \textsl{Ann.\ Math.} \textbf{36} (1935),
607--629.

\bibitem{Whitehead} 
J.\ H.\ C.\ Whitehead, Note on a previous paper entitled `On adding
relations to homotopy groups', {\sl Ann.\ Math.\ }{\bf 47} (1946),
806--810.

J.\ H.\ C.\ Whitehead, Combinatorial homotopy II, {\sl Bull.\ Amer.\
Math.\ Soc.\ }{\bf 55} (1949), 453--496

\bibitem{Witten:1988} 
E.\ Witten, The index of the Dirac operator in loop space, in {\sl
Elliptic Curves and Modular Forms in Algebraic Topology}, ed.\ P.\ S.\
Landweber, Lecture Notes in Mathematics {\bf 1326}, Springer, Berlin,
1988, pp.\ 161--181.

\end{thebibliography}
\end{document}